\documentclass[%
reprint,
superscriptaddress,
 amsmath,amssymb,
 aps,
prb,
floatfix,
]{revtex4-1}

\usepackage{amssymb}
\usepackage{graphicx}
\usepackage{hyperref}
\usepackage{bbold}
\usepackage{braket}
\usepackage{color}

\DeclareMathOperator{\Tr}{Tr}
\DeclareMathOperator{\sgn}{sgn}
\newcommand\numberthis{\addtocounter{equation}{1}\tag{\theequation}}
\newcommand{\mbf}{\mathbf}
\newcommand*{\citen}[1]{%
  \begingroup
    \romannumeral-`\x
    \setcitestyle{numbers}%
    \cite{#1}%
  \endgroup   
}

\definecolor{blue(pigment)}{rgb}{0.2, 0.2, 0.6}
\definecolor{darkerblue}{rgb}{0.0, 0.0, 0.4}
\definecolor{darkblue}{rgb}{0.0,0.0,0.5}
\definecolor{darkgreen}{rgb}{0.0,0.4,0.0}
\hypersetup{
    colorlinks,
    linkcolor=blue(pigment),
    citecolor=darkgreen,
    urlcolor=darkblue
}

\begin{document}

\title{Non-equilibrium Quantum Spin Dynamics \\ from 2PI Functional Integral Techniques in the Schwinger Boson Representation}%

\author{A.\ Schuckert}\email{alexander.schuckert@tum.de}
\affiliation{Institut f\"ur Theoretische Physik, Universit\"at Heidelberg, Philosophenweg 16, 69120 Heidelberg, Germany}
\affiliation{Department  of  Physics,
Technical  University  of  Munich,  85748  Garching,  Germany
}
\author{A.\ \surname{Pi{\~{n}}eiro Orioli}}
\affiliation{Institut f\"ur Theoretische Physik, Universit\"at Heidelberg, Philosophenweg 16, 69120 Heidelberg, Germany}

\author{J.\ Berges}
\affiliation{Institut f\"ur Theoretische Physik, Universit\"at Heidelberg, Philosophenweg 16, 69120 Heidelberg, Germany}
\date{\today}

\begin{abstract}
We present a non-equilibrium quantum field theory approach to the initial-state dynamics of spin models based on two-particle irreducible (2PI) functional integral techniques. It employs a mapping of spins to Schwinger bosons for arbitrary spin interactions and spin lengths. At next-to-leading order (NLO) in an expansion in the number of field components, a wide range of non-perturbative dynamical phenomena are shown to be captured, including relaxation of magnetizations in a 3D long-range interacting system with quenched disorder, different relaxation behaviour on both sides of a quantum phase transition and the crossover from relaxation to arrest of dynamics in a disordered spin chain previously shown to exhibit many-body-localization. Where applicable, we employ alternative state-of-the-art techniques and find rather good agreement with our 2PI NLO results. As our method can handle large system sizes and converges relatively quickly to its thermodynamic limit, it opens the possibility to study those phenomena in higher dimensions in regimes in which no other efficient methods exist. Furthermore, the approach to classical dynamics can be investigated as the spin length is increased.
\end{abstract}

\maketitle

\section{Introduction}

Spin systems are among the most studied models of condensed matter physics owing to their importance in the study of magnetism and related phenomena such as high temperature superconductivity.
Recently, the ability of cold atom experiments to study the initial-state dynamics of spin models and other interacting quantum systems in isolation from the environment has led to the possibility to test many aspects previously not accessible to other experimental platforms.

Understanding how thermal equilibrium emerges from unitary quantum dynamics is one of the most pressing unresolved problems in many-body physics adressable by these experimental platforms. It is conjectured by the eigenstate thermalization hypothesis~\cite{PhysRevA.43.2046,PhysRevLett.80.1373,PhysRevE.50.888,srednicki_approach_1999} that even though the Schr\"odinger equation and its field-theoretic generalizations are time reversible, a general quantum system develops towards a quasi-stationary state which appears to be irreversible, i.e.\ it has effectively forgotten the details about its initial state for relevant observables. This picture has been numerically verified in some models\cite{Rigol2008,eckstein_thermalization_2009,kollath_quench_2007} and studied experimentally in cold atom systems~\cite{trotzky_probing_2012,Newtonscradle,gring_relaxation_2012}. In the context of field theories, thermalization from far-from equilibrium states is quantified by the fulfillment of fluctuation-dissipation relations and has been shown in $O(N)$ symmetric scalar field theories in various spatial dimensions~\cite{Berges2002,BERGES2001369,juchem_quantum_2004} as well as fermionic quantum fields in 3D~\cite{Berges:2002wr} and Heisenberg magnets~\cite{Babadi15} using the same powerful functional integral techniques which we will employ in this paper.

However, it has been discovered that some interacting systems fail to thermalize under the influence of strong disorder, such that the system retains memory of its initial state~\cite{BASKO20061126, Imbrie2016, doi:10.1146/annurev-conmatphys-031214-014726, doi:10.1146/annurev-conmatphys-031214-014701, Schreiber842}. This effect, referred to as many body localization, is currently an active research topic: the dependence on dimensionality~\cite{Bordia} and its stability~\cite{Rubio-Abadal2018} are still topics
of debate. Furthermore, it is of interest to study how the processes underlying thermalization change as aspects of classical mechanics become more important, e.g., by increasing the spin length of a quantum spin system.

So far, the dynamics of spin models have mostly been studied from the point of view of quantum mechanics, where the Schr\"odinger equation is used to evolve the whole many-body wavefunction in order to evaluate the time evolution of observables such as local magnetizations. As for a large quantum system the Hilbert space dimension is too high to allow efficient simulations on a classical computer, truncations such as matrix product states are used to get approximate results, effectively limiting the amount of entanglement which can be captured~\cite{schollwock_density-matrix_2011, verstraete_matrix_2004}. These methods are however mostly restricted to 1D and to early times for thermalizing systems.

In this paper, we offer a different perspective on the dynamics of spin systems in terms of non-equilibrium quantum field theory. Instead of evolving first the state or density operator and from this computing a set of relevant observables, the observables expressed in terms of low-order correlation functions are directly evolved in time. While the corresponding evolution equations are just a reformulation of the 
Schr\"odinger equation, they offer a different route to approximating the quantum dynamics based on two-particle irreducible (2PI) functional integral techniques~\cite{PhysRev.118.1417,Cornwall74,kadbaym}, which are closely related to the Luttinger-Ward formalism~\cite{PhysRev.118.1417,frank_universal_2018,PhysRevB.96.045125}. Motivated by a similar approach in which (pre-)thermalization of a spiral state in a 3D Heisenberg magnet has been shown by mapping spin $1/2$ systems to Majorana fermions~\cite{Babadi15}, we use the Schwinger boson representation to describe the dynamics of spin models with arbitrary spin length $S$ (for a work also employing Schwinger bosons see Ref.~[\citen{PhysRevB.98.045111}]). We employ a non-perturbative approximation based on an expansion in the number of field components to next-to-leading order (NLO)~\cite{Berges2002,Aarts02}, which enables us to also describe strongly interacting systems.

We apply our approach to a range of non-perturbative dynamical phenomena that are known to be challenging and which help demonstrate the characteristic strengths of the functional techniques. Where possible, we employ alternative state-of-the-art methods to benchmark our 2PI NLO results in limiting cases. A particular strength of our method concerns its ability to describe large systems in higher dimensions and to follow the dynamics also to long times. To this end, we first consider relaxation dynamics in a 3D long-range interacting XY spin system with quenched disorder. The problem of the non-equilibrium dynamics of large ensembles of spins with position disorder and interacting
via dipolar interactions is relevant for a number of current experimental realizations ranging from Rydberg atoms~\cite{Signoles17,Barredo1021} to polar molecules~\cite{Yan2013} and NV centers in diamond~\cite{choi_observation_2017}. By solving the evolution equations numerically, we analyze the relaxation dynamics of local magnetizations and unequal-time correlation functions, which
 allow us to describe the effective memory loss of the initial state. 
For bulk quantities, such as the volume-averaged magnetization, we can compare our 2PI NLO results with corresponding results from a diagonalization method applied to sub-clusters of spins (MACE)~\cite{Hazzard14}. We find good agreement when the latter is expected to converge. 

As a further example, we study the relaxation dynamics of a spin chain in a 1D anisotropic XXZ model. In the infinite chain length limit, this model is known to exhibit a quantum phase transition from a gapless Luttinger liquid phase with quasi-long-range order to an (anti-)ferromagnetic phase with long-range order. Computing the time evolution of the staggered magnetization on different sides of the quantum phase transition, we show that our method captures the expected qualitative behavior~\cite{Barmettler2010}. Most remarkably, our results are seen to converge already for rather small system sizes. This illustrates the fast approach of our field theoretic approximation to the thermodynamic limit, such that efficient finite size descriptions can be achieved. These findings open up the possibility to study dynamical
quantum phase transitions in regimes in which other methods such as iMPS~\cite{Barmettler2010} or other DMRG~\cite{RevModPhys.77.259} related methods would fail, e.g.\ in higher dimensions.  

While the first two examples demonstrate the ability of the
Schwinger boson 2PI method to describe thermalization dynamics 
in interacting spin models, the last application concerns 
the dynamical evolution in an interacting system
that refuses to thermalize: a many-body localized (MBL)
system. For this we investigate the paradigmatic example of the 
non-equilibrium dynamics of a Heisenberg spin chain in a random field, initialized in a N\'{e}el ordered state.
Our short-time results indicate a transition from a thermalizing system at
weak disorder, signalized by a vanishing long-time staggered
magnetization, to arrest of the relaxation at strong disorder,
where this quantity is large and nonzero.

This paper is organized as follows. In the first three sections, we develop the Schwinger boson spin-2PI approach, especially trying to make the derivation as transparent as possible for a quick application of the method to other problems. First, we introduce the Schwinger boson representation of spin systems and show how the Schwinger boson constraint is naturally fulfilled in a non-equilibrium quantum field theory formulation. Secondly, we introduce the 2PI effective action and derive the Kadanoff-Baym equations of motion. Thirdly, we employ a non-perturbative approximation to the effective action and show how the resulting approximated Kadanoff-Baym equations can be solved numerically.
In the remaining three sections we apply Schwinger boson spin-2PI to various settings and compare our results with state-of-the art numerical methods.

\section{Non-equilibrium quantum field theory for spin systems}

The aim of this work is to develop a functional integral approach based on the 2PI effective action to describe the non-equilibrium dynamics of quantum spin models using the Schwinger boson representation. Here, we focus on Hamiltonians with couplings $J_{ij}^{\alpha}$ and external fields $B^\alpha_i$ of the type
\begin{equation}
\label{eq:Hgeneral_spins}
\hat{H} = \frac{1}{2}\sum_\alpha \sum_{i\neq j} J_{ij}^{\alpha} \hat{S}^{\alpha}_{i}\hat{S}^{\alpha}_{j} + \sum_\alpha \sum_i B^\alpha_i \hat{S}^{\alpha}_{i},
\end{equation}
where the lower and upper indices denote site and components of the spin operators $\hat{S}^{\alpha}_{i}$, respectively. The $J^{\alpha}_{ij}$ are general, in particular we do not assume nearest-neighbour interactions. The spin operators fulfill the commutation relations
\begin{equation}
\left[\hat{S}_n^\alpha, \hat{S}_m^\beta\right] = i \delta_{nm}\sum_\gamma \epsilon^{\alpha\beta\gamma}\hat{S}^\gamma_n,
\label{eq:spin_commrel}
\end{equation}
and the spin quantum number $S$ is given by
\begin{equation}
\label{eq:spin_length}
\vec{S}^2=S(S+1).
\end{equation}
We note that in comparison to previous works based on a representation in terms of Majorana fermions~\cite{Babadi15,newMajorana}, which is valid for $S=1/2$, our Schwinger boson approach can be applied to arbitrary $S$.

Our first step towards a functional description of quantum spin systems is to derive a path integral formulation. While this procedure is standard for bosonic and fermionic systems~\cite{Peskin95,Altland10}, quantum spin systems are slightly more involved due to their non-trivial commutation relations. One possibility is to use spin coherent states~\cite{Auerbach94}, which leads, however, to topological terms associated to Berry phases. Therefore, a common strategy is to map spins to operators which fulfill canonical algebras and are hence easier to handle. For this, fermions~\cite{PhysRevB.37.3774, PhysRevLett.69.2142}, Holstein-Primakoff bosons~\cite{PhysRev.58.1098, Schmied2010} and even exotic species such as semions~\cite{PhysRevLett.85.5631}, Majorana fermions~\cite{PhysRevB.49.8955} and supersymmetric operators~\cite{PhysRevB.62.3852} have been proposed. In this work, we will employ a Schwinger boson representation, which we introduce and discuss in the following. 

\subsection{Schwinger boson representation}

In the Schwinger boson representation~\cite{Auerbach94}, each spin $\hat S_i^\alpha$ is expressed in terms of two bosons, $\hat a_i$ and $\hat b_i$, via
\begin{align}
\begin{aligned}
\hat{S}_i^x = \frac{1}{2}\left(\hat{b}^{\dagger}_i\hat{a}_i+\hat{a}^{\dagger}_i\hat{b}_i\right)&,\quad
\hat{S}_i^y = \frac{i}{2}\left(\hat{b}^{\dagger}_i\hat{a}_i-\hat{a}^{\dagger}_i\hat{b}_i\right),\\
\hat{S}_i^z = \frac{1}{2}&\left( \hat{a}_i^{\dagger}\hat{a}_i- \hat{b}^{\dagger}_i\hat{b}_i\right),
\label{eq:SB_definition}
\end{aligned}
\end{align}
where the bosonic ladder operators satisfy the algebra $\left[\hat{a}_i, \hat{a}^{\dagger}_j\right]=\left[\hat{b}_i, \hat{b}^{\dagger}_j\right]=\delta_{ij}$, $\Big[\hat{a}_i, \hat{a}_j\Big]=\left[\hat{b}_i, \hat{b}_j\right]=0$, and $\left[\hat{a}_i,\hat{b}_j^{\dagger}\right]=0$. These commutation relations ensure that the mapping (\ref{eq:SB_definition}) fulfills the spin algebra (\ref{eq:spin_commrel}). On top of this, the Schwinger bosons have to fulfill the constraint
\begin{equation}
\hat{n}_i \equiv \hat{a}^{\dagger}_i\hat{a}_i+\hat{b}^{\dagger}_i\hat{b}_i = 2S,
\label{eq:SB_constraint}
\end{equation}
in order to restrict their Hilbert space to the `physical' Hilbert space of the original spins. For instance, for $S=1/2$ the Hilbert space would be comprised of $| 1,0\rangle \leftrightarrow | \!\uparrow \rangle$ and $| 0,1 \rangle \leftrightarrow | \!\downarrow \rangle$. We note that condition (\ref{eq:SB_constraint}) implies (\ref{eq:spin_length}) and is the only place in the Schwinger boson mapping where the spin number $S$ appears. Therefore, the expressions derived in the following sections are valid for arbitrary $S$.

For notational simplicity, it will at times be useful to cast both Schwinger bosons into a two-component complex field operator as
\begin{equation}
	\hat\psi_i^1 \equiv \hat a,\qquad \hat\psi_i^2 \equiv \hat b.
\end{equation}
The commutation relations are then given by $\left[ \hat\psi_i^a , \hat\psi_j^{b\dagger} \right]=\delta_{ij}\delta^{ab}$ and $\left[ \hat\psi_i^a , \hat\psi_j^{b} \right]=0$, and Eq.~(\ref{eq:SB_definition}) can be compactly written as
\begin{equation}
	\hat S_i^\alpha = \frac{1}{2} \hat\psi_i^{a\dagger} \sigma_{ab}^\alpha \hat\psi_i^b ,
\label{eq:SB_trafo_psi}
\end{equation}
where $\sigma^\alpha$, $\alpha\in\{x,y,z\}$, are the Pauli matrices. In this way, the spin Hamiltonian (\ref{eq:Hgeneral_spins}) takes the form
\begin{align}
\label{eq:H_withSB}
\hat{H} =&\, \frac{1}{8}\sum_\alpha \sum_{i\neq j} J_{ij}^{\alpha} \sigma^\alpha_{ab} \sigma^\alpha_{cd}\, \hat\psi_i^{a\dagger} \hat\psi_i^b \hat\psi_j^{c\dagger} \hat\psi_j^d \nonumber\\
&\, + \frac{1}{2} \sum_\alpha \sum_i B^\alpha_i \sigma^\alpha_{ab}\, \hat\psi_i^{a\dagger} \hat\psi_i^b.
\end{align}
Here and in the following, a sum over repeated field component indices $(a,b,c,d)$ is implied, whereas summation over spin component $(\alpha)$ and position $(i,j)$ indices will be explicit.
We note that each term in the above expression is normal ordered, since $i\neq j$.

The validity of the Schwinger boson constraint for suitable approximations will be a major aspect of the discussion in the following sections. In equilibrium, the constraint is usually ensured by introducing a Lagrange multiplier~\cite{Auerbach94}. For non-equilibrium initial value problems, the symmetry-conserving nature of approximations based on the 2PI effective action will automatically conserve the constraint as long as the initial values comply with it. However, in the approximation for the initial state we apply here, only the value of $\langle \hat n_i \rangle$ is explicitly set to the correct value, whereas higher orders of $\hat n_i$ are different. We will discuss ways of improving this limitation in the course of this paper.

\subsection{Functional integral representation}

\begin{figure}[t]
\centering
\includegraphics{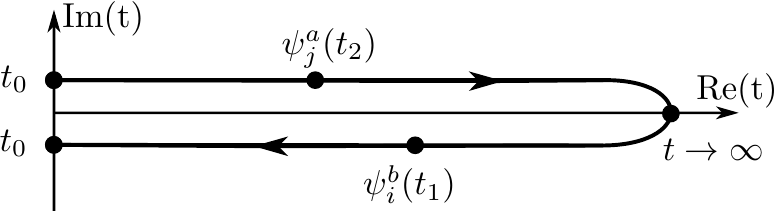}
\caption{In non-equilibrium quantum field theory, all fields are defined on the Schwinger-Keldysh closed time contour. Starting at the initial time $t_0$ it proceeds to infinity ($\mathcal{C}^+$) and then back to the initial time ($\mathcal{C}^-$). Shown is the insertion of two field operators along the contour for the evaluation of a two-point function for the case where the operators are on different branches and $t_2<t_1$.}
\label{fig_closed_contour_2}
\end{figure}
To describe the non-equilibrium dynamics of the above Schwinger boson model we employ its path integral formulation on the Schwinger-Keldysh closed time contour~\cite{Keldysh65} $\mathcal{C}=\mathcal{C}^+\cup\mathcal{C}^-$ depicted in Fig.~\ref{fig_closed_contour_2}, which consists of a forward ($\mathcal{C}^+$) and a backward branch ($\mathcal{C}^-$).
As a first step, we rewrite the identity $Z\equiv \Tr\{ \varrho_0 U(t_0,t) U(t,t_0) \}=1$ as a path integral, where $\varrho_0$ denotes the initial density matrix and the time evolution operator is $U(t,t_0)=\exp\big(-i \hat H (t-t_0)\big)$. For this, we use bosonic coherent states $\ket{\psi}\equiv\bigotimes_i \ket{\psi_i^1}\otimes\ket{\psi_i^2}$, such that $\hat\psi^a_i \ket{\psi} = \psi^a_i \ket{\psi}$.
Following standard procedures~\cite{Berges15,Kamenev2011} leads to
\begin{equation}
\label{eq:Z_functional}
Z = \int \mathrm{d}\psi_0^+\mathrm{d}\psi_0^-\bra{\psi_0^+}\varrho_0\ket{\psi_0^-}\int_{\psi_0^-}^{\psi_0^+}\mathcal{D}'\psi \exp{\left(iS[\bar\psi,\psi]\right)},
\end{equation}
where $\psi_0^+/\psi_0^-$ denote fields on the forward/backward part of the contour at $t=t_0$, and the prime in $\mathcal{D}'\psi$ specifies that integration over $\psi_0^+/\psi_0^-$ is excluded.
The classical action corresponding to model (\ref{eq:H_withSB}) is given by
\begin{align}
	S[\bar\psi,\psi]=&\,\int_\mathcal{C}\mathrm{d}t \Bigg\{\sum_{i}\bar{\psi}_i^a \left( \delta_{ab} i\partial_t - \frac{1}{2} \sum_\alpha B_i^\alpha \sigma_{ab}^\alpha \right) \psi_i^b \nonumber\\
&\qquad\ \ - \frac{1}{8} \sum_\alpha \sum_{i\neq j} J_{ij}^{\alpha} \sigma^\alpha_{ab} \sigma^\alpha_{cd}\, \bar\psi_i^{a} \psi_i^b \bar\psi_j^{c} \psi_j^d \Bigg\} .
\label{eq:action_complex}
\end{align}
The first integral in (\ref{eq:Z_functional}) contains information on the initial state and, as we will see, leads to some complications for spin systems.

\subsection{Local \texorpdfstring{$U(1)$}{U(1)} symmetry and constraints\label{chap_Noethercharge}}

The Schwinger boson mapping, Eq.~(\ref{eq:SB_definition}), is constructed such that raising and lowering operators always appear in pairs. In this way, any spin Hamiltonian written in the Schwinger boson basis, Eq.~(\ref{eq:H_withSB}), conserves the local number of bosons $\hat n_i$, $\left[\hat{H}, \hat{n}_i\right]=0$, and hence fulfills the constraint (\ref{eq:SB_constraint}) at all times.
As a direct consequence of this, the corresponding classical action, Eq.~(\ref{eq:action_complex}), has a local $U(1)$ symmetry parametrized by
\begin{equation}
	\psi_i^a \ \rightarrow \ e^{i\alpha_i} \psi_i^a ,
\label{eq:local_U1_trafo}
\end{equation}
i.e.~both bosons, $\hat a_i$ and $\hat b_i$, are rotated by the same angle.

At the classical level, the local $U(1)$ symmetry leads to a conserved Noether current or continuity equation $\partial_\mu j^\mu_i=0$. Due to the absence of spatial derivatives in the action, Eq.~(\ref{eq:action_complex}), the only non-vanishing component is the temporal one, $j^0_i = - n_i\equiv - \bar\psi_i^a \psi_i^a=-|a_i|^2 - |b_i|^2$. Thus, one obtains \emph{local} number conservation,
\begin{equation}
	\partial_t\, n_i = 0.
\label{eq:Noether}
\end{equation}
The local Schwinger boson constraint is therefore a consequence of the $U(1)$ symmetry of the classical action.

In the full quantum theory, the symmetry of the action leads to a whole hierarchy of Ward-Takahashi identities. Naturally, as we show in Appendix~\ref{app:WTI}, the lowest-order identity is given by
\begin{equation}
	\partial_t \braket{\hat n_i} = 0.
\label{eq:WTI_1}
\end{equation}
Thus, the conservation of $\braket{\hat n_i}$ is directly linked to the local $U(1)$ symmetry of the action. In contrast to gauge theories such as QED or QCD, where a \emph{local} gauge symmetry requires specific terms in the action to cancel each others' contributions, each term in the action (\ref{eq:action_complex}) is individually invariant under a local $U(1)$ transformation. 

Of course, the full operator equation (\ref{eq:SB_constraint}) is not formally equivalent to simply the expectation value equation (\ref{eq:WTI_1}). Instead, Eq.~(\ref{eq:SB_constraint}) implies an infinite hierarchy of identities for expectation values, e.g.~$\braket{\hat n_i^2}=(2S)^2$. As we show in Appendix~\ref{app:WTI}, the conservation of the latter quantity is captured by a second-order Ward-Takahashi identity, namely
\begin{align}
	\partial_t \braket{\hat n_i^2} &= 0.
\label{eq:WTI_2}
\end{align}
Thus, this quantity will fulfill the aforementioned identity provided it is fulfilled at initial time. However, in this work we will consider Gaussian approximations to the initial conditions, which will lead to only $\braket{\hat n_i}$ being set explictly to the right initial value. We will be discussing this approximation, its implications and how it can be overcome in Sec.~\ref{ssec:Gaussian_approx}.

\subsection{Hubbard-Stratonovich transformation}

Approximations for the 2PI effective action become more transparent when dealing with real instead of complex fields, as for example motivated in Ref.~[\citen{Gasenzer2005}]. At the operator level, we split the Schwinger bosons into their real and imaginary parts,
$ \hat{a}_i = (\hat{a}_i^1 + i \hat{a}_i^2)/\sqrt{2}$, $\hat{b} =(\hat{b}_i^1 + i \hat{b}_i^2)/\sqrt{2}$,
where $(\hat{a}_i^{1,2})^\dagger=\hat a_i^{1,2}$ and $(\hat{b}_i^{1,2})^\dagger = \hat b_i^{1,2}$. As before, it is convenient to express these operators in terms of a 4-component real field operator $\hat\varphi_i\equiv\left(\hat a_i^1,\hat a_i^2,\hat b_i^1,\hat b_i^2\right)$.
Inserting this into the Schwinger boson representation (\ref{eq:SB_definition}), the spin operators become
\begin{equation}
	\hat{S}^{\alpha}_i =\frac{1}{4}\hat\varphi^a_i \mathcal{K}_{ab}^{\alpha} \hat\varphi^b_i,
\label{eq:SB_trafo_realphi}
\end{equation}
where
\begin{equation}
	\mathcal{K}^{\alpha}_{ab} = \left[\sigma_x\otimes\mathbb{1}\right]^{ab}\delta^{\alpha x}-\left[\sigma_y\otimes\sigma_y\right]^{ab}\delta^{\alpha y}+\left[\sigma_z\otimes\mathbb{1}\right]^{ab}\delta^{\alpha z} .
\label{eq:S_matrix_def}
\end{equation}
We note that $\mathcal{K}^{\alpha}_{ab}$ is symmetric in the Schwinger boson indices $(ab)$.
Similarly, the equal-time commutation relations can be written as
\begin{equation}
\left[\hat{\varphi}_i^a ,\hat{\varphi}_j^b \right] =-\left[\mathbb{1}\otimes\sigma_y\right]^{ab}\delta_{ij} ,
\label{eq:4_SB_real_comm}
\end{equation}
and the Schwinger boson constraint becomes
\begin{equation}
	\hat{n}_i= \frac{1}{2}\left(\hat{\varphi}^a_i\hat{\varphi}^a_i-2\right)=2S.
\label{eq:SB_constraint_realfields}
\end{equation}

At the level of the path integral, we introduce real fields $\varphi_i^a$ in an analogous way. In doing so, one ought to be careful when comparing identities for operators with those for fields. For instance, while $\hat\psi_i^{a\dagger} \hat\psi_i^a = (\hat{\varphi}^a_i\hat{\varphi}^a_i-2)/2$, see Eq.~(\ref{eq:SB_constraint_realfields}), for fields one finds that $\bar\psi_i^{a} \psi_i^a = \varphi^a_i \varphi^a_i$. Nevertheless, we note that Eqs.~(\ref{eq:SB_trafo_psi}) and (\ref{eq:SB_trafo_realphi}) also hold for fields, i.e.~$\frac{1}{2} \bar\psi_i^{a} \sigma_{ab}^\alpha \psi_i^b = \frac{1}{4}\varphi^a_i \mathcal{K}_{ab}^{\alpha} \varphi^b_i$. Using this, the action in terms of real fields becomes
\begin{align}
	S[\varphi] =&\, \nonumber\\[3pt]
	\int_\mathcal{C}\mathrm{d}t & \Bigg\{ - \frac{1}{2} \sum_i \varphi_i^a \left( \left[\mathbb{1}\otimes\sigma_y\right]^{ab} i\partial_t + \frac{1}{2} \sum_\alpha B_i^\alpha \mathcal{K}_{ab}^\alpha \right) \varphi_i^b \nonumber\\
	&\,\ \  -\frac{1}{2}\sum_\alpha \sum_{i\neq j} J_{ij}^{\alpha} \left(\frac{1}{4}\mathcal{K}_{ab}^{\alpha}\varphi^a_i\varphi_i^b \right) \left(\frac{1}{4}\mathcal{K}_{cd}^{\alpha}\varphi^c_j \varphi_j^d\right) \Bigg\},
\label{eq:action_reals}
\end{align}
where we have discarded boundary terms of the form $\int_\mathcal{C} \partial_t (\varphi_i^1 \varphi_i^2) = 0$.

In order to make the quartic interaction term more tractable, we further introduce an auxiliary (Hubbard-Stratonovich) field $\chi_i^\alpha$ as
\begin{equation}
	\prod_\alpha \int \mathcal{D}\chi^\alpha\, e^{\frac{i}{2} \sum_{ij} \left[ J^{-1} \right]^\alpha_{ij} \chi_i^{\alpha} \chi_j^{\alpha} } = \text{const},
\label{eq:chi_insertion}
\end{equation}
After the substitution $\chi_i^\alpha \rightarrow \chi_i^\alpha - \frac{1}{4} \sum_k J_{ik}^{\alpha}\left( \mathcal{K}_{ab}^{\alpha} \varphi^a_k \varphi^b_k \right)$ the quartic term in (\ref{eq:action_reals}) is replaced by a three-point vertex $\sim \chi\varphi\varphi$ [see Fig.~\ref{vertex}].
Here, we defined $J^\alpha_{ii}=0$ and assumed that the inverse matrix $J^{-1}$ exists, as will be the case in the applications considered in this work. Note that whenever $J^\alpha_{ij}\equiv0$ for a given $\alpha$, the auxiliary field $\chi^\alpha_i$ completely decouples from the $\varphi_i^a$ fields and can hence be ignored. Because of this, in the following, all sums over spin components $\sum_\alpha$ which involve the auxiliary field are to be understood as sums over only those $\alpha$ for which $J^\alpha_{ij}$ does not vanish identically.
Taking this into account, the final action written in terms of $\varphi$ and $\chi$ is given by
\begin{align}
	&S[\varphi,\chi] =\, \notag\\&\,\,
	\int_\mathcal{C}\mathrm{d}t \Bigg\{ - \frac{1}{2} \sum_i \varphi_i^a \left( \left[\mathbb{1}\otimes\sigma_y\right]^{ab} i\partial_t + \frac{1}{2} \sum_\alpha B_i^\alpha \mathcal{K}_{ab}^\alpha \right) \varphi_i^b \nonumber\\
	&\quad+\sum_\alpha \sum_{ij} \left\lbrace\frac{1}{2}\left[J^{-1}\right]_{ij}^{\alpha}\chi_i^{\alpha}\chi_j^{\alpha}-\frac{1}{4} \delta_{ij} \mathcal{K}_{cd}^{\alpha}\, \chi_i^{\alpha} \varphi^c_j\varphi_j^d\right\rbrace \Bigg\}.
\label{eq:action_chi}
\end{align}
With this procedure we have thus rewritten the original spin model in terms of a dynamical 4-component real scalar field $\varphi^a$ and a non-dynamical, in general 3-component real scalar field $\chi^\alpha$.
We note that the coupling factor $J$ has been absorbed into the definition of the auxiliary field, see Eq.~(\ref{eq:chi_insertion}).

\section{2PI generating functional}

\subsection{Generating functional and Gaussian approximation to the initial conditions\label{ssec:Gaussian_approx}}

The starting point to derive the 2PI effective action is to promote $Z$ from Eq.~(\ref{eq:Z_functional}) to a generating functional. For this, we first need to deal with the term $\bra{\psi_0^+}\varrho_0\ket{\psi_0^-}$ related to the initial state.
If the initial state is approximately Gaussian, the density matrix can be parametrized (in the real basis $\varphi$) by~\cite{Berges15}
\begin{equation}
\bra{\psi_0^+}\varrho_0\ket{\psi_0^-} \propto \exp\left(ih_\mathcal{C}[\varphi]\right),
\end{equation}
with
\begin{align}
h_\mathcal{C}[\varphi] =&\, \alpha_0 + \int_{\mathcal{C}}\mathrm{d}t_1\sum_{i}\alpha_{1,i}^a(t_1)\varphi_i^a(t_1) \nonumber \\
&\,+\frac{1}{2!}\int_\mathcal{C}\mathrm{d}t_1\mathrm{d}t_2\sum_{ij}\alpha_{2,ij}^{ab}(t_1,t_2)\varphi_i^a(t_1)\varphi_j^b(t_2) .
\label{eq:IC_gauss_approx}
\end{align}
Here, the functions $\alpha_1$ and $\alpha_2$ only have support at the initial time $t_0$.

In the following, we define a super field $\Phi = (\varphi,\chi)^T$ to contain all $\varphi$ and $\chi$ fields introduced in the previous section. Using Eq.~(\ref{eq:IC_gauss_approx}) we can then promote $Z$ from (\ref{eq:Z_functional}) to the generating functional
\begin{align}
	Z\left[J,R\right] =&\, \int \mathcal{D}\Phi \exp i\left\{ S[\Phi]+\int_\mathcal{C}\mathrm{d}t_1\sum_{i}J_i^a(t_1)\, \Phi_i^a(t_1) \right. \nonumber\\
	&\ \left. + \frac{1}{2}\int_\mathcal{C}\mathrm{d}t_1\, \mathrm{d}t_2 \sum_{ij}\Phi_i^a(t_1)\, R_{ij}^{ab}(t_1,t_2)\, \Phi_j^b(t_2) \right\}.
\label{eq:Zgenerating_functional}
\end{align}
In this expression, the functions $\alpha_1$ and $\alpha_2$ have been absorbed into the sources $J$ and $R$.
Correlation functions can be obtained from the above generating functional by functional derivatives. For example, the first derivatives yield
\begin{align}
	\frac{\delta Z\left[J,R\right]}{i\delta J_{i}^{a}(t)}\bigg|_{J=0=R} =&\, \braket{\Phi_{i}^{a}(t)} \equiv \bar\Phi_{i}^{a}(t),\\
	\frac{\delta Z\left[J,R\right]}{i\delta R_{ij}^{ab}(t_1,t_2)}\bigg|_{J=0=R} =&\, \frac{1}{2} \braket{\mathcal{T}_{\mathcal{C}}\, \Phi_{i}^{a}(t_1) \Phi_{j}^{b}(t_2)} \nonumber\\
	\equiv&\, \frac{1}{2} \left( \mathcal{G}_{ij}^{ab}(t_1,t_2) + \bar\Phi_{i}^{a}(t_1) \bar\Phi_{j}^{b}(t_2) \right) ,
\label{eq:onetwo_functions}
\end{align}
where $\mathcal{T}_{\mathcal{C}}$ is the time ordering operator along the closed time contour $\mathcal{C}$, and we defined the connected two-point correlator $\mathcal{G}$ as well as the field expectation value $\bar\Phi$. We note that $J=0=R$ is a shorthand notation for setting the sources to zero for $t\neq t_0$, whereas for $t=t_0$ one sets $J_i^a \rightarrow \alpha_{1,i}^a$ and $R_{ij}^{ab} \rightarrow \alpha_{2,ij}^{ab}$.

It is important to note that in quantum spin systems, Eq.~(\ref{eq:IC_gauss_approx}) is only an approximation to the correct initial state.
To see this consider a single spin initially in the state $\ket{\uparrow}$. In the Schwinger basis, this corresponds to a Fock state $\ket{1,0}$, which has a non-vanishing connected four-point function,
\begin{align*}
&\braket{1,0|\hat a^{\dagger} \hat a^{\dagger} \hat a \hat a|1,0}_\mathrm{C}\\ &=\braket{1,0|\hat a^{\dagger} \hat a^{\dagger} \hat a \hat a|1,0} - 2 \left(\braket{1,0|\hat a^{\dagger} \hat a|1,0}_\mathrm{C}\right)^2 \\
&= -2, \numberthis
\end{align*}
and is hence non-Gaussian.
Nevertheless, similar to previous related works~\cite{Babadi15}, we will neglect here such non-Gaussian contributions at initial time and approximate the full initial state by the Gaussian form (\ref{eq:IC_gauss_approx}).
One consequence of this will be that, while $\partial_t \braket{\hat n_i^2}=0$ from (\ref{eq:WTI_2}), the identity $\braket{\hat n_i^2}=(2S)^2$ will not be fulfilled at initial time in this approximation (see Appendix~\ref{ChecksAuxCorr}).
While higher-order corrections to (\ref{eq:IC_gauss_approx}) can in principle be added by introducing additional initial time sources~\cite{Garny09}, this is beyond the scope of this work.

\subsection{2PI effective action\label{ch_Effectiveaction}}

The generating functional $Z[J,R]$ is the non-equilbrium quantum field theory generalization of the partition sum in statistical mechanics.
In this sense, the \emph{two-particle-irreducible (2PI) effective action} $\Gamma[\bar\Phi,\mathcal{G}]$ is a free energy analogue defined as the double Legendre transform of the logarithm of $Z[J,R]$ with respect to the source fields $J$ and $R$,
\begin{align*}
\label{3_Gamma2def}
&\Gamma[\bar\Phi,\mathcal{G}]=\log \left(iZ[J,R]\right)-\int_{\mathcal{C}}\mathrm{d}t_1\sum_{i}\bar\Phi_i^a(t_1)J_i^a(t_1)\\
&- \frac{1}{2}\int_{\mathcal{C}}\sum_{ij}\mathrm{d}t_1\mathrm{d}t_2 [\bar\Phi_i^a(t_1)\bar\Phi_j^b(t_2)+\mathcal{G}_{ij}^{ab}(t_1,t_2)]R_{ij}^{ab}(t_1,t_2). \numberthis
\end{align*}
It is parametrized in terms of the field expectation value $\bar\Phi$ and the connected two-point function $\mathcal{G}$. From the above definition one obtains the stationarity conditions
\begin{equation}
\label{eq:eomeffaction}
\frac{\delta \Gamma[\bar\Phi,\mathcal{G}]}{\delta \bar\Phi}\bigg|_{J=0=R}=0, \quad\frac{\delta \Gamma[\bar\Phi,\mathcal{G}]}{\delta \mathcal{G}}\bigg|_{J=0=R}=0,
\end{equation}
which will explicitly be written as equations of motion for $\bar\Phi$ and $\mathcal{G}$ in the next section. These equations further show that $\Gamma[\bar\Phi,\mathcal{G}]$ may be viewed as the quantum generalization of the classical action.

\begin{figure}[t]
\centering
\includegraphics[]{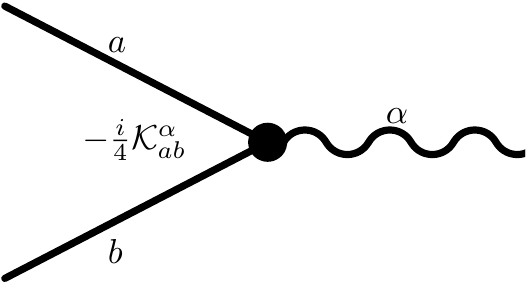}
\caption{The three-point vertex given by the interaction part of the auxiliary field action, where wiggly lines correspond to the auxiliary field correlator $D$ and straight lines to the Schwinger boson correlator $G$. In a loop diagram containing this vertex, all indices would be summed/integrated over, i.e. the Schwinger boson indices $a,b$, the auxiliary field index $\alpha$ as well as the lattice index and time (the latter two are not shown here).}
\label{vertex}
\end{figure}

A very useful decomposition of the 2PI effective action is given by~\cite{Cornwall74}
\begin{equation}
\Gamma[\bar\Phi,\mathcal{G}] = S[\bar\Phi]+\frac{i}{2}\Tr_{\mathcal{C}}\ln \mathcal{G}^{-1} + \frac{i}{2}\Tr_{\mathcal{C}}\left\lbrace \mathcal{G}^{-1}_0[\bar\Phi]\mathcal{G}\right\rbrace +\Gamma_2[\mathcal{G}],
\label{eq:CJT_decomp}
\end{equation}
where a normalization constant was ommited and the free inverse propagator is given by
\begin{equation}
\label{3_freecorr}
i\left[\mathcal{G}_0^{-1}\right]^{ab}_{ij}(t_1,t_2)= \frac{\delta^2S[\bar\Phi]}{\delta\bar\Phi_i^a(t_1)\delta\bar\Phi_j^b(t_2)}.
\end{equation}
The second and third terms in Eq.~(\ref{eq:CJT_decomp}) constitute one-loop quantum corrections to the classical action $S[\bar\Phi]$. The rest functional $\Gamma_2[\mathcal{G}]$ contains the sum of all two-particle-irreducible (2PI) diagrams~\cite{Cornwall74}, made with lines representing the full propagator $\mathcal{G}$ and the interaction vertex [see Fig.~\ref{vertex}]
\begin{equation}
iS_{\mathrm{int}}=-\frac{i}{4} \int_\mathcal{C}\mathrm{d}t\sum_{i,\alpha}\mathcal{K}_{ab}^{\alpha}\, \chi_i^{\alpha}(t)\varphi_i^a(t)\varphi_i^b(t).
\label{eq:int_vertex_2PI}
\end{equation}
Examples of such diagrams will be given in Sec.~\ref{sec:largeN_exp}, where we discuss approximations to $\Gamma_2[\mathcal{G}]$.
That $\Gamma_2[\mathcal{G}]$ is really a sum of 2PI diagrams can be seen by inserting the decomposition (\ref{eq:CJT_decomp}) into the second equation of (\ref{eq:eomeffaction}). One obtains in this way the Schwinger-Dyson equation for the correlator,
\begin{equation}
\label{eq:3_SD_inv}
\mathcal{G}^{-1}=\mathcal{G}_0^{-1}-2i\frac{\delta\Gamma_2[\bar\Phi,\mathcal{G}]}{\delta \mathcal{G}}.
\end{equation}
The last term can be identified with the self-energy, which contains the sum of all 1PI diagrams and therefore $\Gamma_2[\mathcal{G}]$ can only contain 2PI diagrams.

The 2PI effective action constitutes an efficient description of non-equilibrium dynamics, since each diagram in the expansion of $\Gamma_2[\mathcal{G}]$ is built out of the full correlator $\mathcal{G}$, which according to (\ref{eq:3_SD_inv}) already contains an infinite series of diagrams in terms of the bare correlator $\mathcal{G}_0$.
Furthermore, it constitutes a self-consistent description in terms of the physical observables $\bar\Phi$ and $\mathcal{G}$, which does not show secularity problems emerging from expansions in terms of the bare propagator $\mathcal{G}_0$~\cite{Berges2004}.
Finally, because the self-energy is obtained by a functional derivative as in Eq.~(\ref{eq:3_SD_inv}), it is automatically ensured that global conservation laws are fulfilled and that the thermodynamic potentials corresponding to the effective action in thermal equilibrium fulfill all standard relations~\cite{Baym1962}.

\subsection{2PI equations of motion}

In order to rewrite the 2PI equations of motion (\ref{eq:eomeffaction}) and (\ref{eq:3_SD_inv}) in a more convenient form, it is useful to introduce some notation and make some simplifications thanks to the properties of the Schwinger bosons.
We first define the one-point functions
\begin{equation}
	\bar{\chi}_i^{\alpha}\equiv\braket{\chi_i^{\alpha}}, \qquad \bar\varphi_i^a \equiv \braket{\varphi_i^a},
\end{equation}
and the correlators
\begin{align}
G_{ij}^{ab} (t_1,t_2) &= \braket{\mathcal{T}_\mathcal{C}\, \varphi_i^{a}(t_1)\varphi_j^{b}(t_2)} - \bar\varphi_i^{a}(t_1)\bar\varphi_j^{b}(t_2),
\label{eq:G_def}\\
M_{ij}^{\alpha b} (t_1,t_2) &= \braket{\mathcal{T}_\mathcal{C}\, \chi_i^{\alpha}(t_1)\varphi_j^{b}(t_2)} - \bar\chi_i^{\alpha}(t_1)\bar\varphi_j^{b}(t_2), \\
D_{ij}^{\alpha\beta} (t_1,t_2) &= \braket{\mathcal{T}_\mathcal{C}\, \chi_i^{\alpha}(t_1)\chi_j^{\beta}(t_2)}-\bar{\chi}_i^{\alpha}(t_1)\bar{\chi}_j^{\beta}(t_2), \label{4_def_D}
\end{align}
Due to the Schwinger boson constraint (\ref{eq:SB_constraint}), it turns out that both $\bar\varphi \equiv 0$ and $M \equiv 0$. This can most easily be seen by noting that within the Hilbert space allowed by (\ref{eq:SB_constraint}), any expectation value of an uneven number of Schwinger boson fields must be zero. The construction of the Schwinger bosons ensures that initially $\bar\varphi$ and $M$ vanish. In this case, the 2PI equations derived in the next section show that these quantities will remain zero throughout the evolution, regardless of the approximation made. However in general $\bar\chi\neq 0$ because $\chi\sim\varphi\varphi$.

Due to these simplifications, the free inverse propagator $\mathcal{G}^{-1}_{0}$ becomes
\begin{equation}
\mathcal{G}^{-1}_{0,ij}\left(t_1,t_2\right) = 
-i\begin{pmatrix}
\frac{\delta^2S[\bar\varphi,\bar{\chi}]}{\delta\bar\varphi_i^{a}(t_1)\delta\bar\varphi_j^{b}(t_2)} & 0 \\
0 & \frac{\delta^2S[\bar\varphi,\bar{\chi}]}{\delta\bar{\chi}_i^{\alpha}(t_1)\delta\bar{\chi}_j^{\beta}(t_2)}
\end{pmatrix},
\end{equation}
with components
\begin{align}
	\frac{\delta^2S[\bar\varphi,\bar{\chi}]}{\delta\bar\varphi_i^{a}(t_1)\delta\bar\varphi_j^{b}(t_2)} =&\, -\delta_\mathcal{C}(t_1-t_2)\bigg\lbrace \left[ \mathbb{1}\otimes\sigma_y\right]^{ab} i\partial_{t_1}  \nonumber\\ 
	+ \frac{1}{2}\sum_\alpha & \left( \bar{\chi}_j^{\alpha}(t_1) + B_j^\alpha \right)\mathcal{K}_{ab}^{\alpha}\bigg\rbrace \delta_{ij},
\label{4_G_0}\\
	\frac{\delta^2S[\bar\varphi,\bar{\chi}]}{\delta\bar{\chi}_i^{\alpha}(t_1)\delta\bar{\chi}_j^{\beta}(t_2)} =&\, \left[J^{-1}\right]_{ij}^{\alpha} \delta^{\alpha\beta} \delta_\mathcal{C}(t_1-t_2).
\end{align}
Similarly, the full correlation function can be written as
\begin{equation}
\mathcal{G}=
\begin{pmatrix}
G&0 \\
0&D
\end{pmatrix},
\label{eq:prop_structure}
\end{equation}
and the self-energy as
\begin{equation}
2i\frac{\delta\Gamma_2[\mathcal{G}]}{\delta\mathcal{G}} \equiv \begin{pmatrix}
\Sigma & 0 \\
0& \Pi
\end{pmatrix}.
\label{4_selfenergies_def}
\end{equation}

We are now ready to derive the equations of motion from the 2PI effective action as given by Eq.~(\ref{eq:eomeffaction}) for the non-vanishing one- and two-point functions. The equation for the auxiliary field expectation value $\bar\chi$ follows from $\delta\Gamma/\delta \bar{\chi} |_{J=0=R} = 0$ and is given by
\begin{align}
	\bar{\chi}^{\alpha}_i(t)&=\frac{1}{4}\sum_j J_{ij}^{\alpha}\mathcal{K}^{\alpha}_{cd}G^{cd}_{jj}(t,t) = \sum_j J_{ij}^{\alpha} \braket{\hat S^{\alpha}_j(t)},
\label{Chi_eom}
\end{align}
where the last equality follows from the definition of $G$, Eq.~(\ref{eq:G_def}), and Eq.~(\ref{eq:SB_trafo_realphi}) [see Sec.~\ref{sct_summary} for further details].
Similarly, the equations for the correlators $G$ and $D$ can be obtained from (\ref{eq:3_SD_inv}) by convoluting it with $\mathcal G$ from the right to obtain
\begin{widetext}
\begin{align}
\left(\left[\mathbb{1}\otimes\sigma_y\right]^{ac} i\partial_{t_1}+\frac{1}{2}\sum_{\alpha} \left( \bar{\chi}_i^{\alpha}(t_1) + B_i^\alpha \right)\mathcal{K}_{ac}^{\alpha}\right)G_{ij}^{cb}&(t_1,t_2)=-i\delta^{ab}\delta_{ij}\delta_\mathcal{C}(t_1-t_2)-i\int_\mathcal{C}\mathrm{d}t \sum_k\Sigma_{ik}^{ac}(t_1,t)G_{kj}^{cb}(t,t_2)
\label{4_KB_G}, \\
D_{ij}^{\alpha\beta}&(t_1,t_2)=iJ_{ij}^{\alpha}\delta^{\alpha\beta}\delta_\mathcal{C}(t_1-t_2)+i\sum_k J_{ik}^{\alpha}\int_\mathcal{C}\mathrm{d}t\sum_{l, \delta} \Pi_{kl}^{\alpha\delta}(t_1,t)D_{lj}^{\delta\beta}(t,t_2)
\label{4_KB_D}.
\end{align}
\end{widetext}
The left hand side of Eq.~(\ref{4_KB_G}) shows that $\bar\chi$ acts as an effective external field for the correlator $G$. The above equations are the Kadanoff-Baym or 2PI equations of motion for the Schwinger boson and auxiliary field correlators. The first is linked to magnetizations and, as we show in appendix \ref{Chapter_SpinObsfromAux}, the latter to spin correlators. Without further approximation, these two equations simply constitute a reformulation of the Schr\"odinger equation for these two observables. In practice, one must however employ approximations to the self-energies $\Pi$, $\Sigma$, which we motivate and employ in the next section.

\section{Non-perturbative expansion\label{sec:largeN_exp}}

\subsection{\texorpdfstring{$1/N$}{1/N} expansion to NLO}

Applications of the 2PI effective action to non-equilibrium problems such as thermalization require approximations of the functional $\Gamma_2$ beyond leading order (LO) to include direct scatterings. When a small interaction parameter is available, perturbative or loop approximations can yield accurate results~\cite{BERGES2001369, juchem_quantum_2004,MBLKnap}. A powerful non-perturbative method for $N$-component field theories consists in expanding $\Gamma_2$ in powers of $1/N$~\cite{Berges2002,Aarts02}. When taking into account diagrams up to next-to-leading (NLO) order, this approximation has been shown to outperform other beyond-mean-field approximation schemes in ultracold Fermi~\cite{Kronenwett2011} and Bose gases~\cite{Gasenzer2009,Gasenzer2005,Branschadel}, including optical lattices~\cite{Rey04,Temme2006}, and it has been successfully applied to a myriad of problems such as thermalization of bosonic~\cite{BERGES2001369} and fermionic quantum fields~\cite{Berges:2002wr}, time evolution of quasi-particle spectral functions~\cite{Aarts:2001qa}, critical exponents in the $O(N)$ model~\cite{Alford2004}, prethermalization and heating in Floquet systems~\cite{Weidinger2017}, and the Kondo effect in the Anderson impurity model~\cite{Bock2016}. Remarkably, it has even been able to capture regimes of very large infrared fluctuations close to nonthermal fixed points~\cite{Orioli:2015dxa,Berges:2017ldx,walz_large-n_2017,Berges2017}, as well as regimes of strong couplings even for rather small values of $N$~\cite{Berges2017}. Note however that the dynamics of topological defects have been shown not to be reproduced by this approximation using a homogeneous background field~\cite{berges_topological_2011,rajantie_looking_2006,rajantie_counting_2010}.

As shown in Sec.~\ref{chap_Noethercharge}, our Schwinger boson theory has a local $U(1)$ symmetry, which corresponds to a local $O(2)$ symmetry in the basis of real fields $\varphi$. After the Hubbard-Stratonovich transformation the action (\ref{eq:action_chi}) still has the same symmetry and the $\chi$ field does not participate in the transformation. Thus, in our case we will perform a $1/N$ expansion with $N=2$ analogously to the above examples.
In order to do so, we need to classify the 2PI diagrams contributing to $\Gamma_2$ in terms of their scaling with $N$. This requires the identification of all $O(N)$ invariants~\cite{Aarts02} that can arise due to the interaction vertex (\ref{eq:int_vertex_2PI}) and the propagator structure (\ref{eq:prop_structure}). The possible $O(N)$ invariants are given by
\begin{equation}
	\Tr\big\{ (\mathcal{K} G)^n \big\}\quad \text{and} \quad D,
\label{eq:ON_invariants}
\end{equation}
where the trace is taken over the field component indices. 
That $D$ is an invariant can be seen from the fact that $\chi$ does not participate in the $O(N)$ transformation. To see why $\Tr\big\{ (\mathcal{K} G)^n \big\}$ is an invariant as well, note first that the combination $\mathcal{K}^\alpha_{ab}\varphi_i^a \varphi_i^b$ is invariant under $O(N=2$), since it is equivalent to $2 \sigma_{ab}^\alpha \bar\psi_i^{a} \psi_i^b$, which is $U(1)$ invariant with Eq.~(\ref{eq:local_U1_trafo}). Therefore, $\Tr\big\{ \mathcal{K} G \big\}$ is an $O(N)$ invariant. The generalization to $\Tr\big\{ (\mathcal{K} G)^n \big\}$ is then straightforward, since such a term arises from $n$ copies of $\mathcal{K}^\alpha_{ab}\varphi_i^a \varphi_i^b$.

Given the $O(N)$ invariants of (\ref{eq:ON_invariants}) the next step is to establish their scaling with $N$. Due to the trace operation we have
\begin{equation}
	\Tr\big\{ (\mathcal{K} G)^n \big\} \sim N.
\end{equation}
To find out the scaling of $D$ we first note that the generalization of our action (\ref{eq:action_chi}) to an $O(N)$ symmetric theory for general $N$ requires the renormalization of the coupling as $J_{ij} \rightarrow J_{ij}/N$ in order for the $N\rightarrow\infty$ limit to exist. Taking this into account, it then follows from Eq.~(\ref{4_KB_D}) that $D$ must scale as
\begin{equation}
	D \sim \frac{1}{N}.
\end{equation}

\begin{figure}[t]
\centering
\vspace*{2ex}
\includegraphics{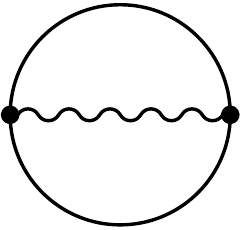}
\caption{The only diagram contributing at NLO, which is also the only loop diagram with two vertices. Feynman rules are the same as in Fig.~\ref{vertex}.}
\label{fig_NLO_diag}
\end{figure}
The leading-order (LO) approximation to the 2PI effective action is given by setting
\begin{equation}
	\Gamma_2^{\mathrm{LO}}=0,
\label{eq:Gamma2_LO}
\end{equation} 
and taking only the one-loop corrections of Eq.~(\ref{eq:CJT_decomp}) into account. This corresponds to taking into account contributions $\sim N$ as can be seen from inserting the free Schwinger boson correlator (\ref{4_G_0}) into the third term, yielding a term $\sim \Tr{\lbrace\mathcal{K} G\rbrace}$ in the effective action (which reappears on the LHS of Eq.~(\ref{4_KB_G})). A contribution to the equation of motion for $D$ coming from this term does not appear to this order as it is $\sim N^0$. This means that $D=0$ to LO, so that to this order the connected spin correlators vanish (Appendix~\ref{Chapter_SpinObsfromAux}). Equivalently, this corresponds to a Hartree-Fock approximation in the theory with only $\varphi$ fields, Eq.~(\ref{eq:action_reals}).

The only next-to-leading-order (NLO) contribution to the effective action in powers of $N$ is given by the diagram in Fig.~\ref{fig_NLO_diag}, which scales as $\sim N^0$ and corresponds to
\begin{widetext}
\begin{equation}
\Gamma_2^{\mathrm{NLO}}=-2i\left(\frac{-i}{4}\right)^2 \frac{1}{2!}\sum_{\alpha\beta}\mathcal{K}_{ab}^{\alpha}\mathcal{K}_{cd}^{\beta}\int_\mathcal{C}\mathrm{d}t_1\int_\mathcal{C}\mathrm{d}t_2\sum_{jk}G^{ac}_{jk}(t_1,t_2)G^{bd}_{jk}(t_1,t_2)D_{jk}^{\alpha\beta}(t_1,t_2).
\label{eq:Gamma2_NLO}
\end{equation} 
\end{widetext}
Here, an overall $(-i)$ due to the definition of $\Gamma_2$ is included, as well as a $1/2!$ factor from the expansion of the exponential and a combinatorial factor of $2$. In this work, we will employ the NLO approximation, Eq.~(\ref{eq:Gamma2_NLO}), and neglect higher-order contributions in $1/N$.
Of course, in our case $N=2$, so $1/N$ is not a particularly small number. Nevertheless, even for small $N$ the $1/N$ expansion to NLO has been shown to yield surprisingly good results in a variety of systems where other known approaches fail~\cite{Alford2004,Berges2002}, and it has been successfully applied in related works on quantum spin dynamics using a Majorana representation~\cite{Babadi15, newMajorana}.

It is important to note that, while the diagram of Fig.~\ref{fig_NLO_diag} also corresponds to the lowest non-vanishing contribution to $\Gamma_2$ in a perturbative expansion in the $(\varphi,\chi)$-basis, this apparent equivalence is lifted at NNLO~\cite{Aarts2002}. The non-perturbative nature of Eq.~(\ref{eq:Gamma2_NLO}) can be best seen by integrating out $\chi$, which yields an infinite series of diagrams to arbitrary order in the coupling~\cite{Aarts2002}. This demonstrates the power of the auxiliary field formalism which manages to encapsulate large sets of corrections into $\chi$.

In the rest of the section, we discuss the Schwinger boson constraint after truncation of $\Gamma_2$, derive the equations of motion following from the LO and NLO approximations, discuss the initial conditions and give a brief summary of the whole method including a mapping to observables in the original spin basis.

\subsection{Schwinger boson constraint in \texorpdfstring{$2$}{2}PI approximations \label{SBconstr_approx}}

As we mentioned in Sec.~\ref{chap_Noethercharge} (c.f.~App.~\ref{app:WTI}), the conservation of the set of identities $\braket{(\hat n_i)^k}=(2S)^k, k \in \mathbb{N}$, following from the Schwinger boson constraint (\ref{eq:SB_constraint}) is directly associated in the 2PI formalism to the local $U(1)$ symmetry of the action. The latter implies the Ward-Takahashi identities $\partial_t \braket{(\hat n_i)^k}=0$, which ensures the fulfillment of the above identities if they are fulfilled at $t=0$. Thus, the identities $\braket{\hat n_i^k}=(2S)^k$ will be true in our 2PI approximation as well if the truncation of $\Gamma_2$ preserves the local $U(1)$ symmetry and the associated Ward-Takahashi identities. Note that in the case of the Gaussian initial state we use here, only the $k=1$ identity is explicitly fulfilled at $t=0$ as argued in Sec.~\ref{ssec:Gaussian_approx}.

Each term of the action (\ref{eq:action_chi}) is separately invariant under the local $U(1)$ symmetry (see Sec.~\ref{chap_Noethercharge}). As a consequence, the vertex (\ref{eq:int_vertex_2PI}), which constitutes the building block of all diagrams contributing to $\Gamma_2$, is invariant as well. This can be shown explicitly by inspecting
\begin{equation}
	-\frac{i}{4} \int_\mathcal{C}\mathrm{d}t\sum_{i,\alpha}\mathcal{K}_{ab}^{\alpha}\, D_{ij_1}^{\alpha\delta_1}(t,t_1) G_{ij_2}^{ad_2}(t,t_2) G_{ij_3}^{bd_3}(t,t_3),
\end{equation}
which constitutes the functional representation of the vertex (\ref{eq:int_vertex_2PI}). In the above expression, the variables $t_{1\ldots3}$, $j_{1\ldots3}$, $\delta_1$ and $d_{2,3}$ are free and would be connected to other vertices in a full diagram contributing to the effective action. Following similar arguments as above for the $O(N)$ invariants, namely that $\mathcal{K}^\alpha_{ab}\varphi_i^a \varphi_i^b \sim \sigma_{ab}^\alpha \bar\psi_i^{a} \psi_i^b$, this expression is invariant under the local $U(1)$ or $O(2)$ symmetry, which means that each diagram in $\Gamma_2$ is \emph{individually} invariant. This is in contrast to QED, where only a subset of diagrams taken together are invariant under the U(1) gauge symmetry \cite{Reinosa07}. 

In conclusion, approximations in 2PI, and in particular the $1/N$ expansion to NLO introduced above, respect the local $U(1)$ symmetry of action (\ref{eq:action_chi}) and the associated Ward-Takahashi identities. Thus, the Schwinger boson identity $\braket{\hat n_i}=2S$ will be respected by our approximations. Higher-order Ward-Takahashi identities will also be respected, but the associated conserved quantities may not have the appropriate initial values due to the Gaussian approximation of Sec.~\ref{ssec:Gaussian_approx} as mentioned before (see App.~\ref{app:WTI} for details).

\subsection{LO equations of motion}\label{LO_approx}

At leading order, Eq.~(\ref{eq:Gamma2_LO}), the self-energies are zero, $\Sigma=\Pi=0$, and so $D$ is trivial according to Eq.~(\ref{4_KB_D}). At this order, the only interaction effect is due to the $\bar{\chi}$ term on the left hand side of the equation of motion for the correlator $G$, Eq.~(\ref{4_KB_G}). We will show in the following that this equation is equivalent to the mean field equations of motion for the spin expectation values, also known as Bloch equations.

For practical purposes, it is useful to express the correlator in terms of functions that are not time-ordered along the closed time contour. Thus, we decompose $G$ in spectral (commutator) and statistical (anticommutator) components as~\cite{Berges15}
\begin{equation}
\label{decomp_F_Rho}
	G^{ab}_{ij}(t_1,t_2) = F^{ab}_{ij}(t_1,t_2) - \frac{i}{2}\sgn_\mathcal{C}(t_1-t_2) \rho^{ab}_{ij}(t_1,t_2) ,
\end{equation}
where $F$ and $\rho$ are given by
\begin{align}
	F^{ab}_{ij}(t_1,t_2) =&\, \frac{1}{2}\braket{\left\lbrace \hat\varphi_i^a(t_1), \hat\varphi_j^b(t_2)\right\rbrace}_c,\\
	\rho^{ab}_{ij}(t_1,t_2) =&\, i\braket{\left[ \hat\varphi_i^a(t_1), \hat\varphi_j^b(t_2)\right]}.
\end{align}
The subscript ``$c$'' indicates the connected correlator. Inserting this into (\ref{4_KB_G}) with $\Sigma=\Pi=0$, using that $\partial_{t_1}\sgn_\mathcal{C}(t_1-t_2) = 2\delta_\mathcal{C}(t_1-t_2)$  and employing the commutation relations (\ref{eq:4_SB_real_comm}) results in decoupled equations for $F$ and $\rho$, 
\begin{align}
\partial_{t_1}F_{ij}^{ab}(t_1,t_2)&= \nonumber\\
	\frac{i}{2}\sum_{\gamma}\big(\bar{\chi}_i^{\gamma}&(t_1)+B_i^\gamma\big) \left[\mathbb{1}\otimes\sigma_y\right]^{ac}\mathcal{K}_{cd}^{\gamma}F_{ij}^{db}(t_1,t_2),
\label{eq:Feq_LO}\\
\partial_{t_1}\rho_{ij}^{ab}(t_1,t_2)&= \nonumber\\
	\frac{i}{2}\sum_{\gamma} \big(\bar{\chi}_i^{\gamma}&(t_1)+B_i^\gamma\big) \left[\mathbb{1}\otimes\sigma_y\right]^{ac}\mathcal{K}_{cd}^{\gamma}\rho_{ij}^{db}(t_1,t_2).
\end{align}

To map these equations onto equations for spin variables, we first note that
\begin{equation}
\label{mag_S}
\braket{\hat S^{\alpha}_i(t)} = \frac{1}{4}\mathcal{K}_{ab}^{\alpha} F^{ab}_{ii} (t,t),
\end{equation}
which follows from $\sgn_\mathcal{C}(0)=0$ and Eq.~(\ref{eq:SB_trafo_realphi}). Moreover, we employ
\begin{equation}
\mathcal{K}_{ac}^{\alpha}\mathcal{K}_{bd}^{\beta}\left[\mathbb{1}\otimes\sigma_y\right]^{dc}=i \sum_\gamma \epsilon^{\alpha\beta\gamma}\mathcal{K}_{ab}^{\gamma},\label{S_matrix_identity}
\end{equation}
which we prove in App.~\ref{app:Kidentity}. 
Using Eq.~(\ref{eq:Feq_LO}), and Eq.~(\ref{Chi_eom}) for $\bar\chi$, we thus arrive at
\begin{align*}
&\partial_t \braket{\hat S^{\alpha}_i (t)} \\
&\quad= \frac{i}{4}\mathcal{K}_{ac}^{\alpha}\left[\mathbb{1}\otimes\sigma_y\right]^{cd}\sum_\beta\mathcal{K}_{bd}^{\beta} \big( \bar{\chi}_i^{\beta}(t) + B_i^\beta \big) F_{ii}^{ba}(t,t) \\
&\quad= \sum_{\beta\gamma} \epsilon^{\alpha\beta\gamma} \big( \bar{\chi}_i^{\beta}(t) + B_i^\beta \big) \frac{1}{4} \mathcal{K}_{ab}^{\gamma}F_{ii}^{ab}(t,t) \\
&\quad= \sum_{\beta\gamma} \epsilon^{\alpha\beta\gamma} \bigg( \sum_j J_{ij}^\beta \braket{S_j^{\beta}(t)} + B_i^\beta \bigg) \braket{S_i^{\gamma}(t)} ,\numberthis
\end{align*}
which constitute the mean field (Bloch) equations for the Hamiltonian (\ref{eq:Hgeneral_spins}).

\subsection{NLO equations of motion}

At next-to-leading order, Eq.~(\ref{eq:Gamma2_NLO}), the self-energies follow from (\ref{4_selfenergies_def}) as
\begin{align}
	&\Sigma_{ij}^{\mathrm{NLO},ab}(t_1,t_2) = 2i \frac{\delta \Gamma_2}{\delta G_{ij}^{ab}(t_1,t_2)}\nonumber \\
	&\qquad= -\frac{1}{4}\sum_{\alpha\beta}\mathcal{K}_{ac}^{\alpha}\mathcal{K}_{bd}^{\beta}G_{ij}^{cd}(t_1,t_2)D_{ij}^{\alpha\beta}(t_1,t_2),
\label{4_G_SE}\\
	&\Pi_{ij}^{\mathrm{NLO},\alpha\beta}(t_1,t_2) = 2i \frac{\delta \Gamma_2}{\delta D_{ij}^{\alpha\beta}(t_1,t_2)} \nonumber\\
	&\qquad=-\frac{1}{8}\mathcal{K}_{ab}^{\alpha}\mathcal{K}_{cd}^{\beta}G^{ac}_{ij}(t_1,t_2)G^{bd}_{ij}(t_1,t_2) .
\label{4_D_SE}
\end{align}
Similar to the correlator $G$ in Eq.~(\ref{decomp_F_Rho}), it is convenient to split the self-energies into spectral and statistical parts as
\begin{align}
	\Pi^{\alpha\beta}_{ij}(t_1,t_2) =&\, \Pi^{F,\alpha\beta}_{ij}(t_1,t_2) - \frac{i}{2}\sgn_\mathcal{C}(t_1-t_2) \Pi^{\rho,\alpha\beta}_{ij}(t_1,t_2), \\
	\Sigma^{ab}_{ij}(t_1,t_2) =&\, i\delta_\mathcal{C}(t_1-t_2)\,\Sigma^{(0),ab}_{ij}(t_1) \nonumber\\
	+&\, \Sigma^{F,ab}_{ij}(t_1,t_2) - \frac{i}{2}\sgn_\mathcal{C}(t_1-t_2) \Sigma^{\rho,ab}_{ij}(t_1,t_2) ,
\end{align}
where we separated a possible time-local part $\Sigma^{(0)}$~\cite{Aarts02}. In the same way, we decompose the correlator $D$ into
\begin{equation}
D_{ij}^{\alpha\beta}(t_1,t_2)= iJ_{ij}^{\alpha}\delta^{\alpha\beta}\delta_\mathcal{C}(t_1-t_2)+\sum_{kl} J_{ik}^{\alpha}\hat{D}_{kl}^{\alpha\beta}(t_1,t_2)J_{lj}^{\beta},\label{4_D_decomp}
\end{equation}
and define
\begin{equation}
\label{decomp_DF_Drho}
\hat D^{\alpha\beta}_{ij}(t_1,t_2) = \hat D^{F,\alpha\beta}_{ij}(t_1,t_2) - \frac{i}{2}\sgn_\mathcal{C}(t_1-t_2) \hat D^{\rho,\alpha\beta}_{ij}(t_1,t_2) .
\end{equation}

The NLO equations of motion, following from inserting Eqs.~(\ref{4_G_SE}) and (\ref{4_D_SE}) with the above decompositions into Eqs.~(\ref{4_KB_G}) and (\ref{4_KB_D}), can be greatly simplified using the properties of the Schwinger bosons. As shown in Ref.~[\citen{Babadi15}] and also Sec.~\ref{init_state}, physical initial states imply that $G_{ij}(t_0,t_0)\sim\delta_{ij}$ and hence $\Sigma_{ij}(t_0,t_0)\sim\delta_{ij}$. This property extends to all later times by induction through Eqs.~(\ref{4_KB_G}) and (\ref{4_KB_D}). Because of this, we can replace
\begin{align}
G^{ab}_{ij}(t_1,t_2)\ &\rightarrow\ \delta_{ij}\, G_{ii}^{ab}(t_1,t_2), \\
\Sigma^{ab}_{ij}(t_1,t_2)\ &\rightarrow\ \delta_{ij}\, \Sigma_{ii}^{ab}(t_1,t_2).
\end{align}
This means in particular that the local part of the self energy, $\Sigma^{(0)}$, vanishes,
\begin{equation}
\Sigma^{(0)}_{ii} \sim J_{ii} = 0.
\end{equation}
At NLO one further finds from (\ref{4_D_SE}) that also
\begin{equation}
\Pi^{\alpha\beta}_{ij}(t_1,t_2)\rightarrow\delta_{ij}\Pi_{ii}^{\alpha\beta}(t_1,t_2).
\end{equation}

All in all, the 2PI equations for the Schwinger boson and auxiliary field correlators become
\begin{widetext}
\begin{align}
	\partial_{t_1}F_{ii}^{ab}(t_1,t_2)&= i\left[\mathbb{1}\otimes\sigma_y\right]^{ac}\bigg\lbrace\frac{1}{2} \sum_{\gamma} \big( \bar{\chi}_i^{\gamma}(t_1) + B_i^\gamma \big) \mathcal{K}_{cd}^{\gamma}F_{ii}^{db}(t_1,t_2) \nonumber\\
	&\,\qquad\qquad\qquad\ +\int_0^{t_1}\mathrm{d}t\,\Sigma_{ii}^{\rho,cd}(t_1,t)F_{ii}^{db}(t,t_2)-\int_0^{t_2}\mathrm{d}t\,\Sigma^{F,cd}_{ii}(t_1,t)\rho^{db}_{ii}(t,t_2)\bigg\rbrace ,
\label{eq:eomF_NLO}\\
	\partial_{t_1}\rho_{ii}^{ab}(t_1,t_2)&= i\left[\mathbb{1}\otimes\sigma_y\right]^{ac}\bigg\lbrace\frac{1}{2}\sum_{\gamma} \big( \bar{\chi}_i^{\gamma}(t_1) + B_i^\gamma \big) \mathcal{K}_{cd}^{\gamma}\rho_{ii}^{db}(t_1,t_2)+\int_{t_2}^{t_1}\mathrm{d}t\, \Sigma_{ii}^{\rho,cd}(t_1,t)\rho_{ii}^{db}(t,t_2)\bigg\rbrace,
\label{eq:eomrho_NLO}
\end{align}
and
\begin{align}
	\hat{D}^{F,\alpha\beta}_{kj}(t_1,t_2) &= -\Pi_{kk}^{F,\alpha\beta}(t_1,t_2)\delta_{kj} + \int_0^{t_1} \mathrm{d}t \sum_{m,\delta} \Pi_{kk}^{\rho, \alpha\delta}(t_1,t)J_{km}^{\delta}\hat{D}^{F,\delta\beta}_{mj}(t,t_2) - \int_0^{t_2} \mathrm{d}t \sum_{m,\delta} \Pi_{kk}^{F,\alpha\delta}(t_1,t)J_{km}^{\delta}\hat{D}^{\rho,\delta\beta}_{mj}(t,t_2),
\label{eq:eomDF_NLO}\\
	\hat{D}^{\rho,\alpha\beta}_{kj}(t_1,t_2) &= -\Pi_{kk}^{\rho,\alpha\beta}(t_1,t_2)\delta_{kj} + \int_{t_2}^{t_1}\mathrm{d}t\,\sum_{m,\delta}  \Pi_{kk}^{\rho,\alpha\delta}(t_1,t)J_{km}^{\delta}\hat{D}^{\rho,\delta\beta}_{mj}(t,t_2).
\label{eq:eomDrho_NLO}
\end{align}
\end{widetext}
Note that the above equations are simply a reformulation of Eqs.~(\ref{4_KB_G}) and (\ref{4_KB_D}) in terms of spectral and statistical components, which removed the reference to a closed time contour.

The approximation to NLO enters through the spectral and statistical components of the self-energies, which are given by
\begin{align}
&\Sigma^{F,ab}_{ii}(t_1,t_2) = -\frac{1}{4}\sum_{\alpha\beta}\mathcal{K}_{ac}^{\alpha}\mathcal{K}_{bd}^{\beta}\sum_{kl}J_{ik}^{\alpha}J_{li}^{\beta}\nonumber\\&\,\times\bigg(F_{ii}^{cd}(t_1,t_2)\hat{D}_{kl}^{F,\alpha\beta}(t_1,t_2) -\frac{1}{4}\rho_{ii}^{cd}(t_1,t_2)\hat{D}^{\rho,\alpha\beta}_{kl}(t_1,t_2)\bigg), \label{4_Sigma_F}\\ 
&\Sigma^{\rho,ab}_{ii}(t_1,t_2) = -\frac{1}{4}\sum_{\alpha\beta}\mathcal{K}_{ac}^{\alpha}\mathcal{K}_{bd}^{\beta}\sum_{kl}J_{ik}^{\alpha}J_{li}^{\beta}\nonumber\\&\,\times\bigg(\rho_{ii}^{cd}(t_1,t_2)\hat{D}_{kl}^{F,\alpha\beta}(t_1,t_2) +F_{ii}^{cd}(t_1,t_2)\hat{D}^{\rho,\alpha\beta}_{kl}(t_1,t_2)\bigg),\label{4_Sigma_rho}
\end{align}
and
\begin{align}
\label{4_Pi_F}
	\Pi^{F,\alpha\beta}_{ii}(t_1,t_2) &= -\frac{1}{8}\mathcal{K}_{ab}^{\alpha}\mathcal{K}_{cd}^{\beta}\bigg(F^{ac}_{ii}(t_1,t_2)F^{bd}_{ii}(t_1,t_2)\nonumber\\
	&\qquad-\frac{1}{4}\rho^{ac}_{ii}(t_1,t_2)\rho^{bd}_{ii}(t_1,t_2)\bigg), \\
\label{4_Pi_rho}
	\Pi^{\rho,\alpha\beta}_{ii}(t_1,t_2) &= -\frac{1}{4}\mathcal{K}_{ab}^{\alpha}\mathcal{K}_{cd}^{\beta}F^{ac}_{ii}(t_1,t_2)\rho^{bd}_{ii}(t_1,t_2).
\end{align}
Apart from this, the auxiliary field one-point function is still given by Eq.~(\ref{Chi_eom}), i.e.~
\begin{equation}
\label{eq:eomChi_NLO}
\bar{\chi}^{\alpha}_j(t)=\frac{1}{4}\sum_k J_{jk}^{\alpha}\mathcal{K}^{\alpha}_{cd}F^{cd}_{kk}(t,t).
\end{equation}
The above equations can be further simplified assuming spatially homogeneous initial states and fields as well as translationally  invariant interactions,  see App.~\ref{Init_state_hom} for details.

\subsection{Initial conditions\label{init_state}}

The 2PI equations derived in the previous sections are first-order integro-differential equations which can be solved numerically by providing them with initial conditions. In our case, we need to provide only $F_{ii}^{ab}(t_0,t_0)$ and $\rho_{ii}^{ab}(t_0,t_0)$, since the initial values for $\hat D^F$ and $\hat D^\rho$ follow directly from evaluating the right hand side of Eqs.~(\ref{eq:eomDF_NLO}) and (\ref{eq:eomDrho_NLO}) at the initial time.

The initial conditions for $F$ are related to the initial values of the magnetizations $\hat S_i^\alpha$ and the spin quantum number $S$. This can be seen by writing $F$ in the complex basis and then using (\ref{eq:SB_definition}) and the constraint (\ref{eq:SB_constraint_realfields}) to relate it to the spin observables. For instance, one has at initial time $F^{11}=\frac{1}{2}\braket{\lbrace\hat{a}_1,\hat{a}_1\rbrace} = \braket{\hat{a}^{\dagger}\hat{a}}+\frac{1}{2}=\braket{\hat{S}^z}+S+\frac{1}{2}$. In this way, one obtains
\begin{widetext}
\begin{align}
F_{ii}(t_0,t_0)
=\begin{pmatrix}
\braket{\hat{S}_i^z}+S+\frac{1}{2} & 0 & \braket{\hat{S}_i^x} &\braket{\hat{S}_i^y} \\
0 &\braket{\hat{S}_i^z}+S+\frac{1}{2}& -\braket{\hat{S}_i^y} &\braket{\hat{S}_i^x}\\
\braket{\hat{S}_i^x} &-\braket{\hat{S}_i^y}& -\braket{\hat{S}_i^z}+S+\frac{1}{2} &0\\
\braket{\hat{S}_i^y} &\braket{\hat{S}_i^x}& 0 & -\braket{\hat{S}_i^z}+S+\frac{1}{2}
\end{pmatrix},
\label{eq:IC_F}
\end{align}
\end{widetext}
where all spin expectation values are evaluated at the initial time $t_0$. We note that this is the only point in our theory where the length of the spin appears. By setting $S$ in the initial condition above, the length of the spin is fixed for the whole time evolution as the identity $\braket{\hat n_i}=2S$ is conserved. We point out that in the Gaussian approximation to the initial conditions considered here, all correlators of the form $\braket{\hat{S}^{\alpha}_i(0)\hat{S}^{\beta}_j(0)}_c$ with $i\neq j$ are initially zero as is shown in appendix \ref{ChecksAuxCorr}. This is indeed sufficient for the initial product states we will consider in this work, which have no initial correlations.

The initial conditions for $\rho_{ii}^{ab}$ are determined by the equal-time commutation relations given in Eq.~(\ref{eq:4_SB_real_comm}). Writing the matrix out explicitly, they are given by
\begin{align*}
\rho_{ii}(t_0,t_0)&= \begin{pmatrix}
0 & -1 & 0 & 0 \\
1&0&0&0 \\
0&0&0&-1\\
0&0&1&0
\end{pmatrix}.\numberthis
\label{eq:IC_rho}
\end{align*}
\section{Summary of the method\label{sct_summary}}

In summary, we have developed a Schwinger boson 2PI description of quantum spin models of type (\ref{eq:Hgeneral_spins}) using a formulation in terms of a 4-component real scalar field $\varphi^a$ and a (in general) 3-component real scalar (auxiliary) field $\chi^\alpha$. In this basis, we have performed a $1/N$ expansion to NLO by making use of the local $O(2)$ symmetry associated to the Schwinger boson constraint (\ref{eq:SB_constraint}). The resulting equations of motion for the average field $\bar\chi$ and the correlators $F$, $\rho$, $D^F$, and $D^\rho$ are given in Eqs.~(\ref{eq:eomF_NLO}), (\ref{eq:eomrho_NLO}), (\ref{eq:eomDF_NLO}), (\ref{eq:eomDrho_NLO}), and (\ref{eq:eomChi_NLO}), which are complemented by the initial conditions given in Eqs.~(\ref{eq:IC_F}) and (\ref{eq:IC_rho}). In the case of homogeneous initial conditions and translationally invariant interactions, the equations can be simplified as given in App.~\ref{Init_state_hom}. Details on the numerical procedure can be found in App.~\ref{Num_impl}.

Various spin observables can be extracted out of the $\varphi$ and $\chi$ field correlators. The one-point function, or site-resolved magnetization, can be obtained in two equivalent ways via
\begin{align}
	\braket{\hat S^{\alpha}_i(t)} &= \frac{1}{4}\mathcal{K}_{ab}^{\alpha} F^{ab}_{ii} (t,t) \\
	&= \sum_k \left[J^{-1}\right]_{ik}^{\alpha} \bar\chi_k^{\alpha}(t),
\label{eq:spin1pt_phichi}
\end{align}
as demonstrated above and in App.~\ref{Chapter_SpinObsfromAux}. In the latter appendix we further show that the correlator $D$ can be related to the connected two-point functions of the spin variables via
\begin{align}
	\frac{1}{2} \big\langle \big\{ \hat S_i^\alpha(t_1) , \hat S_j^\beta(t_2) \big\} \big\rangle_c &= \hat{D}_{ij}^{F,\alpha\beta}(t_1,t_2),
\label{eq:spin2ptF_phichi}\\
	i \big\langle \big[ \hat S_i^\alpha(t_1) , \hat S_j^\beta(t_2) \big] \big\rangle &= \hat{D}_{ij}^{\rho,\alpha\beta}(t_1,t_2).
\label{eq:spin2ptrho_phichi}
\end{align}
It is important to note that these relations only strictly hold in the exact theory. An alternative way to compute two-point spin functions, which correspond to 4-point functions in the $\varphi$ fields, would be via a Bethe-Salpeter type equation as done in Ref.~[\citen{Babadi15}].

In the remaining sections we will apply the developed formalism to a variety of problems by solving the 2PI equations of motion numerically. For this we use a predictor-corrector algorithm and check that the Schwinger boson identity $\braket{\hat n_i}=2S$ is fulfilled~(see App.~\ref{Num_cons}). By comparing it to different state-of-the art numerical techniques we will show that it is able to capture most important aspects of the thermalization processes present in various different setups.

\section{Magnetization dynamics of a 3D dipolar interacting spin system with quenched disorder\label{ch_rel_qu}}
In this section, we apply Schwinger boson spin-2PI to the non-equilibrium dynamics of large ensembles of spins with positional disorder and dipolar interactions. Such systems are relevant for a number of current experimental realizations ranging from Rydberg atoms~\cite{Signoles17,Barredo1021} to polar molecules~\cite{Yan2013} and NV centers in diamond~\cite{choi_observation_2017}. Specifically, we consider a dipolar XY model with Hamiltonian
\begin{equation}
\label{XY_model}
\hat{H} = \frac{1}{2}\sum_{i\neq j}J_{ij}\left(\hat{S}^x_i\hat{S}^x_j+\hat{S}^y_i\hat{S}^y_j\right)+\Omega\sum_i\hat{S}^x_i,
\end{equation}
and interactions given by
\begin{equation}
\label{dipolarJij}
J_{ij} = \frac{C_3 (1-\cos^2(\theta_{ij}))}{|\vec{r}_i-\vec{r}_j|^3}.
\end{equation}
In the above expression, $\vec{r}_i$ is the position of atom $i$ and $\theta_{ij}$ is the angle between $\vec{r}_i-\vec{r}_j$ and the quantisation axis.
We consider the relaxation dynamics of a relatively large system of 100 spins in three spatial dimensions starting in an initial product state given by
\begin{equation}
\label{initial_state}
\ket{\Psi_0} = \otimes_{i} \ket{\downarrow}_i.
\end{equation}
Such a system size is well beyond reach for calculations based on exact diagonalization and beyond the scopes of DMRG due to the high dimensionality.

The parameters we choose for the model in Eq.~(\ref{XY_model}) are motivated by the Rydberg experiment of Ref.~[\citen{Signoles17}] with interaction strength $C_3/\hbar = -2\pi\times1.73\,\mathrm{GHz}\,\mu\mathrm{m}^3$ and Rabi frequency $\Omega/\hbar= 2\pi\times 1.48\,\mathrm{MHz}$. Similar to that experiment, we consider a three-dimensional cloud of spins with random positions taken from a Gaussian distribution and impose a low-distance cut-off due to the Rydberg blockade~\cite{Comparat:10}. The parameters are again taken from the above reference.

Because of this setting, the $J_{ij}$ are inhomogeneous but constant over each realization of the system, i.e.\ there is quenched disorder. Due to the $1/r^3$ dependence of the interactions, a few entries of $J_{ij}$ are rather large, and lead to slow convergence of the time step in the numerics. Since we are mainly interested in a benchmark of the 2PI method, we set all entries above $J_\text{cut}/\hbar=3\times 10^6\,\mathrm{Hz} $ to exactly $J_\text{cut}$. This is similar in spirit to soft core potentials such as those realized with Rydberg dressing~\cite{PhysRevX.4.041037}.

The inhomogeneity of the system drastically increases the numerical cost of solving 2PI equations, as compared to translation-invariant systems, see App.~\ref{Init_state_hom} and Ref.~[\citen{Babadi15}]. We note, however, that the 2PI approach still only scales polynomially with system size and furthermore converges comparably quickly to its thermodynamic limit as discussed in Sec.~\ref{Rel_dyn_XXZ}.

\begin{figure}[b]
\centering
\includegraphics{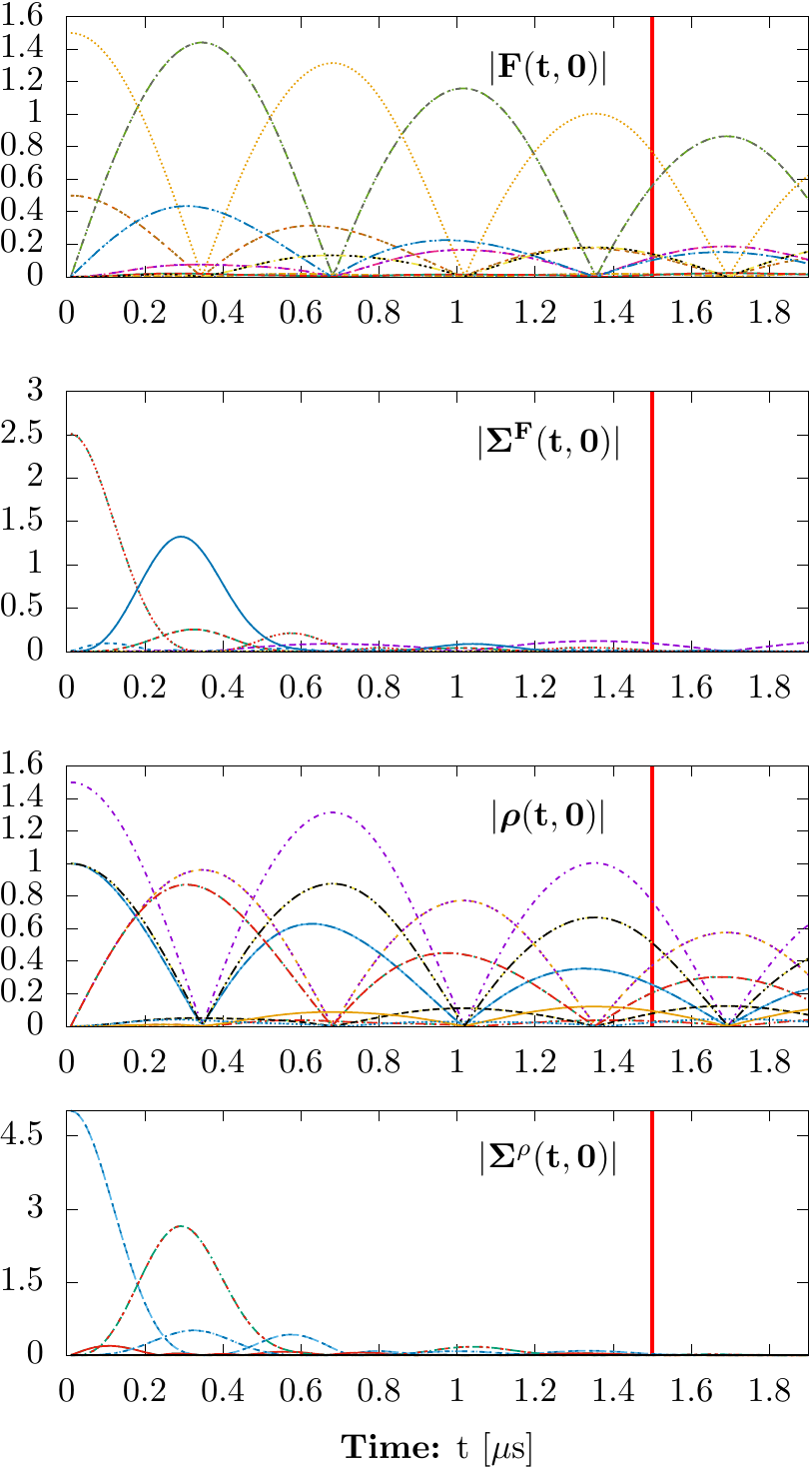}
\caption{The correlations with the initial time in the dipolar XY Hamiltonian with $100$ randomly placed spins starting in a $-z$ polarized state (same system as in Fig.~\ref{fig_mag_N100_2}) as quantified by all $16$ components of the Schwinger boson self-energy $\Sigma^{F/\rho}(t,0)$ and correlators $F(t,0)/\rho(t,0)$ at one site (similar graphs are obtained at all sites). The red line indicates the memory zone, which is shifted along with the time evolution. The damping in all of these functions reflects the effective loss of memory of the initial state as expected for a thermalizing system.}
\label{fig_memcut}
\end{figure}
\subsection*{Memory cut and computational resources}

The memory integrals of the 2PI equations, which integrate from the initial to the current time, require storage of all past times to compute the next step. Since the two-point correlators $G$ and $D$ depend on two time variables, this implies that the memory requirements scale quadratically with the number of time steps. This is further worsened by the fact that $D$ depends independently on two space indices due to the inhomogeneity. All in all, this makes it a priori difficult to evolve the system to long times.

However, the contributions from early times to time-evolving observables such as correlation functions effectively become less important at later times. In our theory, this effective loss of memory can be quantified by the dynamics of the self-energies $\Sigma^{F/\rho}(t,0)$ and the Schwinger boson correlators $F(t,0)$, $\rho(t,0)$, both of which appear in the memory integrals. As a two-time function with argument $(t,0)$ measures the correlation with the initial state, we expect it to approach zero as $t\rightarrow \infty$, at least for a thermalizing system.

In Fig.~\ref{fig_memcut} we show the early time dynamics of these functions for the $100$ spin system to be studied below. In all cases, we see a clear damping of the envelope of the oscillations with time. In particular, the self-energies are shown to decrease by at least a factor of $10$ within the time window shown. This suggests that contributions to the memory integrals from the distant past will be strongly suppressed and hence can be safely ignored.
This enables us to restrict the memory integrals to only the recent history, which significantly reduces the amount of resources needed.

For the simulations presented below we choose to restrict the memory of the system to around $1.5\mu s$, which is marked as a red line in Fig.~\ref{fig_memcut}. We have tested that variations in the memory cut chosen do not significantly affect the results presented, see App.~\ref{ch_ressources} for a more detailed description of this procedure. Moreover, in this section and the next we show results for one realization of the disorder only, but have tested for early times that averages over five realizations give similar results. This indicates that in this description sufficient self-averaging occurs already for these relatively small system sizes.

\subsection*{Demagnetization dynamics}

\begin{figure}[htp]
\centering
\includegraphics{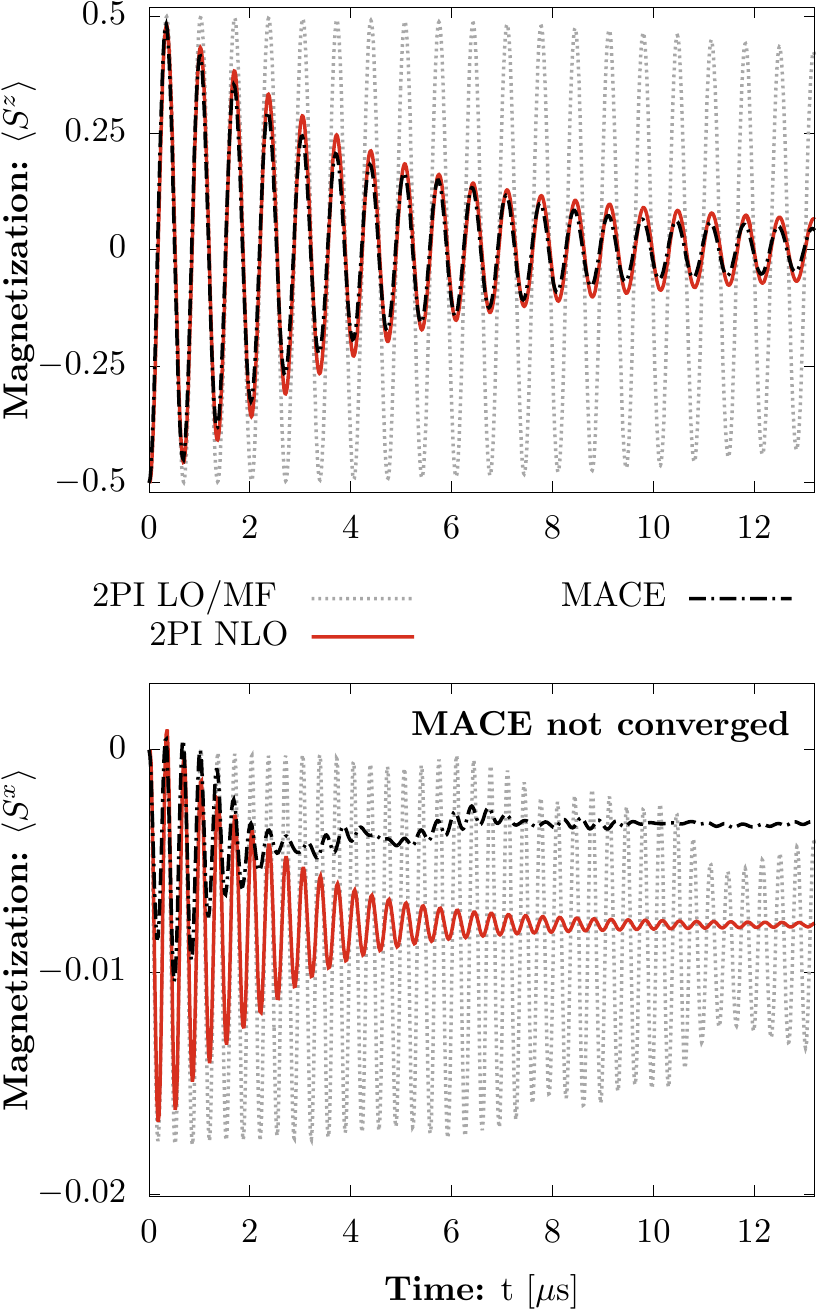}
\caption{Time evolution of the magnetizations in the dipolar XY Hamiltonian with $100$ randomly placed spins starting in a $-z$ polarized state. MACE, mean field (2PI LO/MF) and $2$PI NLO are compared. The $S^y$ component shows similar behaviour as the $S^z$ component, shifted by a quarter period of the oscillation. We only show a single realization of the quenched disorder here, but have checked for small times that an averaged result shows qualitatively similar behaviour. We emphasize that the MACE result for the $S^x$ component has not converged yet as shown in App.~\ref{Num_impl}.}
\label{fig_mag_N100_2}
\end{figure}

We compute the relaxation dynamics of the volume-averaged magnetization
\begin{equation}
\braket{\hat S^\alpha}= \frac{1}{N}\sum_{i=1}^N \braket{\hat S_i^\alpha}
\end{equation}
for the model and initial conditions specified above. To test the accuracy of the Schwinger boson spin-2PI method, we compare the results to the cluster method MACE~\cite{Hazzard14}, which is well-suited for studying the early-time dynamics of one-point functions in such disordered systems and has shown to work well in various similar long-range interacting spin models~\cite{Yan2013,Hazzard14,Signoles17}. For more details on this method see appendix \ref{MACE}.
Apart from the effective loss of memory of the initial state, thermalization is characterized by the approach of observables to their corresponding equilibrium values. The time evolution of the $\alpha = x,z$ magnetization components are shown in Fig.~\ref{fig_mag_N100_2}. Starting from the initial state given in Eq.~(\ref{initial_state}), which is an eigenstate of the XY Hamiltonian, the $S^z$ component (and equally the $S^y$ component not shown) starts to oscillate at a period $T= 2\pi/\Omega \approx 0.67\,\mu\mathrm{s}$ determined by the Rabi frequency. Due to interactions, these oscillations decay in time and they approach zero, as shown by MACE (black dashed line). This behavior can be understood from the fact that the Hamiltonian does not favour any particular direction along $z$. Hence, one would expect $\braket{\hat S^z}=0$ in equilibrium as long as the final equilibration temperature is above possible symmetry-breaking transitions..

From a dynamical point of view, at a mean-field level, the inhomogeneity of the interactions causes each spin to oscillate at a different effective frequency, which leads to dephasing of the total magnetization. The build-up of correlations beyond mean-field leads to additional damping of the magnetization~\cite{Signoles17}, as can be seen in Fig.~\ref{fig_mag_N100_2}.

In the LO or mean-field approximation (gray dotted line), the damping is extremely slow. In fact, it is present only due to the inhomogeneity of the system, as was also noted in Ref.~[\citen{Signoles17}] and will be seen in the next sections. In the limit where the system becomes (discretely) translational invariant, e.g.~on a lattice, the mean field approximation would lead to no damping, therefore failing to describe thermalization in this closed quantum system. 
The failure of the LO approximation to describe the relaxation dynamics comes as no surprise since it does not account for direct scattering effects~\cite{Berges2002}.

The NLO approximation (red thick line), on the other hand, reproduces the damping of $\braket{\hat S^z}$ shown by MACE remarkably well.
As the damping rate is related to the imaginary part of the self energy, we expect quantitative agreement to improve in higher orders of the $1/N$ expansion.
Although the expected equilibrium value of vanishing magnetization in the $z$ and $y$ components is not reached in the simulated time span, the monotonically damped oscillations around zero are a strong indicator for a relaxation to this value.
In contrast to the $z$-component of the magnetization, the MACE and NLO curves for $\braket{\hat S^x}$ show no agreement between each other, even at early times. For this particular observable, however, MACE has not reached convergence yet for the maximal cluster size employed here, namely $13$, as shown in App.~\ref{MACE}. Thus, the MACE prediction shown for $\braket{\hat S^x}$ can not be taken as a quantitatively accurate result to compare with.
\subsection*{Energy}
\begin{figure}[htp]
\centering
\includegraphics{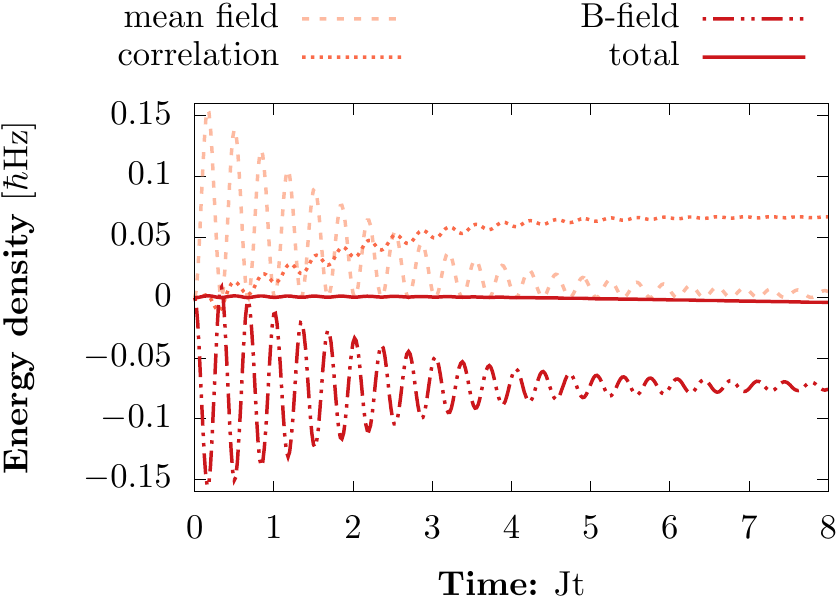}
\caption{Different contributions to the total energy as calculated via the formulae derived in App.~\ref{app:energy}. The mean field part of the energy slowly dampens while the contribution of the quantum correlations rises. The energy stored in the B-field contribution approaches a finite value due to the finite $S^x$ magnetization for long times. The total energy stays constant as expected from the symmetry conserving properties of the approximation.}
\label{fig_energy}
\end{figure}

As a complementary characterisation of the dynamics, we display in Fig. \ref{fig_energy} the time evolution of different contributions to the energy, namely: the mean-field or disconnected part $J\braket{S}\braket{S}$, the linear 'B-field' component $B\braket{S}$ and the connected contributions arising from quantum correlations $J\braket{SS}_c$. We give the expression of the energy along with a description of its derivation in App.~\ref{app:energy}. Due to the conserving properties of our approximation, the total energy stays constant up to small numerical errors. It furthermore vanishes for the fully $-z$ polarized initial product state considered here.

In general, the different energy components show clear oscillations at twice the Rabi frequency. Every time the total spin crosses the $xy$-plane the interaction energy rises and is correspondingly compensated by a negative B-field energy contribution. Initially, the interaction energy is just given by the mean-field part but as time passes correlations build up and the interaction energy becomes dominated by the connected part of the $\braket{SS}$ correlator. At long times, the correlation energy saturates and compensates the negative energy contribution coming from the residual total $S_x$ magnetization. This shows once again the importance of fluctuations beyond mean-field in the long-time dynamics of the system.

To summarize, we have efficiently simulated a system of 100 spins in 3D governed by a dipolar XY model with quenched disorder in an external field using the Schwinger boson spin-2PI method. Taking advantage of the loss of memory from the initial state, we were able to simulate the dynamics to relatively long times, despite the memory requirements imposed by the inhomogeneity of the problem.
In a regime where mean-field clearly fails to describe the relaxation process, the NLO result for the total $\braket{\hat S^z}$ magnetization agrees remarkably well with the behavior predicted by MACE. While considerable deviations are observed for the $\braket{S^x}$ component, which are at least partly attributed to a lack of convergence of MACE, both NLO and MACE predict a non-vanishing long-time value for this observable. Thus, these results show the potential of the present method to describe the relaxation dynamics of spin systems of considerable size ($\gtrsim100$) in high dimensions up to relevant thermalization time scales, even for inhomogeneous problems.
\section{Relaxation dynamics around the quantum phase transition of the anisotropic XXZ chain \label{Rel_dyn_XXZ}}
The question of whether and how the far-from-equilibrium dynamics on different sides of a quantum critical point (QCP) are connected to the underlying equilibrium quantum phase transition~\cite{Barmettler2010,PhysRevB.97.174401} has recently gained much attention from the perspective of dynamical phase transitions~\cite{heyl_dynamical_2014,PhysRevLett.119.080501,PhysRevLett.115.140602,PhysRevLett.110.135704,PhysRevB.96.104436}. In this section, we investigate whether this field of study may be addressed by our 2PI method, similarly to what has been done for an $O(N)$ model in Ref.~[\citen{PhysRevB.96.134313}]. Here, we consider a model studied before in this context~\cite{Barmettler2010,heyl_dynamical_2014}, the antiferromagnetic nearest-neighbor interacting XXZ chain with periodic boundary conditions defined by the Hamiltonian
\begin{equation}
	\hat{H}= J \sum_{i} \left(\hat{S}_i^x\hat{S}_{i+1}^x+ \hat{S}_i^y\hat{S}_{i+1}^y+\Delta\hat{S}_i^z\hat{S}_{i+1}^z\right),
\label{eq:H_anisotropicXXZ}
\end{equation}
where we choose $J>0$ and $\Delta$ denotes the anisotropy.
This model exhibits an equilibrium quantum phase transition from a gapless Luttinger liquid phase with quasi-long-range order for $|\Delta|<1$, to an antiferromagnetic (ferromagnetic) phase with long-range order for $\Delta >1$ ($\Delta<-1$)~\cite{Barmettler2010}.

We study the evolution of the staggered magnetization,
\begin{equation}
	\sum_i (-1)^i \braket{S_i^z(t)},
\label{eq:stag_magn_def}
\end{equation}
in a spin chain initialized with classical N\'{e}el order, i.e.
\begin{equation}
	\ket{\Psi_0} = \ket{\uparrow\downarrow\uparrow\cdots\uparrow\downarrow},
\label{eq:IC_Neel}
\end{equation}
for different anisotropies $\Delta$.
The time evolution of this initial state has been extensively studied with a numerically accurate method (iMPS) in the infinite length limit~\cite{Barmettler2010, PhysRevLett.102.130603}. Those studies show different dynamical behaviour of this non-equilibrium initial state depending on $\Delta$. One finds exponentially damped oscillations with near constant oscillation period for $\Delta\leq 1$, a simple exponential decay for $\Delta>1$, and an algebraic decay for $\Delta=0$. This behaviour has later been attributed to an underlying dynamical quantum phase transition (DQPT) at $\Delta=1$~\cite{heyl_dynamical_2014}, with the long-time average of the staggered magnetization being the order parameter of the transition.

\begin{figure}[htp]
\centering
\includegraphics{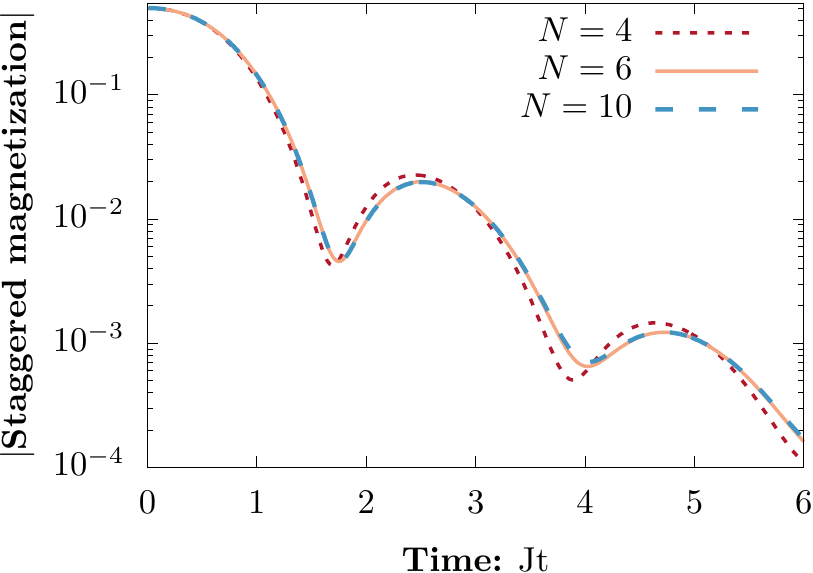}
\caption{Comparison of different system sizes in the dynamics of the staggered magnetization starting in the classical N\'{e}el state in the XXZ chain at the Heisenberg point $\Delta=1.0$. A fast convergence to the thermodynamic limit is found as chain length $6$ shows no sizeable difference to chain length $10$.}
\label{fig_finite_size_flow}
\end{figure}

\subsection*{Evaluation in the infinite length limit}

Using our 2PI approach, we first compare in Fig.~\ref{fig_finite_size_flow} the effect of varying the system size on the dynamics of the staggered magnetization for $\Delta=1$. Remarkably, we find no significant changes in the dynamics for the times considered when increasing the chain length from $N=6$ to $N=10$. This suggests that results with a system size of just $N=6$ can already be taken as a good approximation to the thermodynamic limit in this particular problem.
Moreover, we observed a similarly fast convergence to the thermodynamic limit in two spatial dimensions (not shown), which indicates that this method is also well-suited for the study of quantum dynamics of spin systems in the infinite volume limit in higher dimensions. This fast convergence to the thermodynamic limit is a feature resulting from the field-theoretic nature of our method and was also found in Ref.~[\citen{Babadi15}].
We note that, in contrast to the previous section, we do not use a memory cut here as we found it to lead to an unphysical leveling-off of the exponential damping.

\begin{figure}[htp]
\centering
\includegraphics{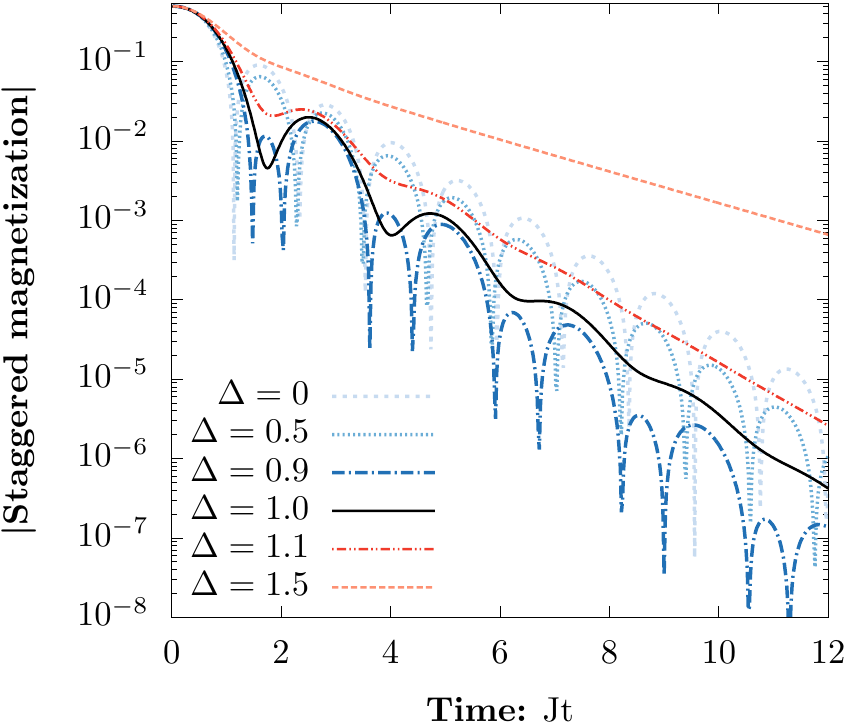}
\caption{Time evolution of the modulus of the staggered magnetization in the XXZ chain with different anisotropies $\Delta$, tuning across the (dynamical) quantum phase transition at $\Delta=1$. The dynamics exhibit an oscillating exponentially damped behaviour for $\Delta<1$ and a pure exponential damping for $\Delta>1$, which was previously found with MPS~\cite{Barmettler2010}. Note that in mean field/LO the staggered magnetization stays constant for all times and hence all features seen here are solely obtained from the NLO approximation.}
\label{fig_rel_HB_NN}
\end{figure}
\begin{figure}[htp]
\centering
\includegraphics{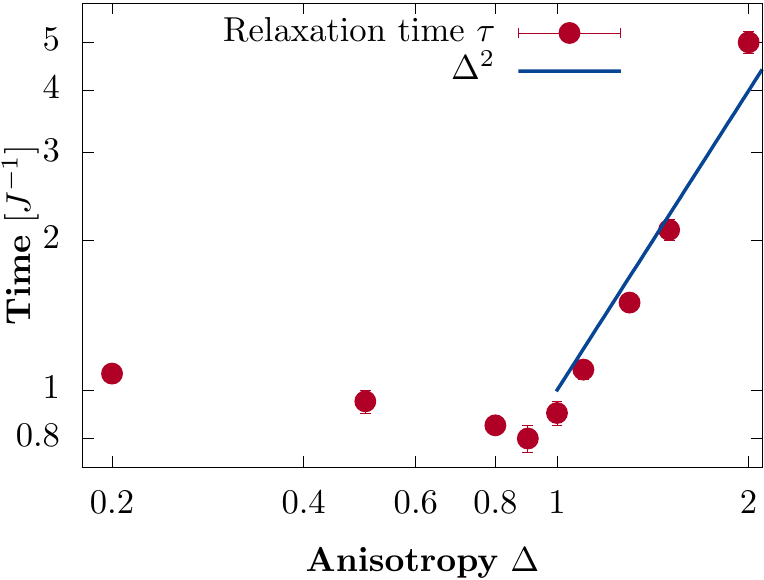}
\caption{Relaxation time as obtained from an exponential fit to the data in Fig.~\ref{fig_rel_HB_NN} as a function of anisotropy. As the QCP around $\Delta=1$ is approached the dynamics becomes faster, which is the anomalous behaviour found before in this model. Furthermore, the relaxation time changes asymetrically as the QCP is approached from above/below. The errors result from the fitting procedure and are smaller than the dot size for most data points. The blue line (not a fit) indicates the $\Delta^2$ behaviour previously found for the approach to the QCP from above\cite{Barmettler2010}.}
\label{fig_reltimes_HB}
\end{figure}

\subsection*{Dynamics of the N\'{e}el ordered state on different sides of the QCP}

Fig. \ref{fig_rel_HB_NN} shows the time evolution of the staggered magnetization for different values of $\Delta$ below and above the transition as obtained from our 2PI approximation. Remarkably, our method captures the qualitative behavior expected~\cite{Barmettler2010}. For $\Delta<1$ we obtain exponentially damped oscillations, whereas for $\Delta>1$ the damping becomes exponential and non-oscillatory.
This represents a considerable improvement compared to previous mean-field treatments based on a mapping to a spinless fermion model~\cite{Barmettler2010}, which found spurious algebraic decay of the staggered magnetization for $\Delta<1$ and a constant oscillatory behaviour for $\Delta>1$.
Note that such a mean-field approximation does not correspond to our LO approximation, which is equivalent to mean-field in the original spin variables and which does not show \emph{any} dynamics here.

Fitting an exponentially damped function $f(t) \sim \exp(-t/\tau)$ to our data, where the proportionality factor contains an oscillatory function for $\Delta<1$, we extract the relaxation time $\tau$ as a function of the anisotropy, see Fig.~\ref{fig_reltimes_HB}. As the critical point at $\Delta=1$ is approached from below we observe a fall-off of the relaxation time, which is the behaviour expected in this model. Note that this constitutes a rather anomalous behavior compared to the usual critical slowing down close to quantum critical points~\cite{Zinn-Justin:572813,Barmettler2010}. As $\Delta=1$ is approached from above, an algebraic dependence $\tau(\Delta) \sim \Delta^2$ has been previously found in Ref.~[\citen{Barmettler2010}]. Fig.~\ref{fig_reltimes_HB} shows that our results are compatible with such a quadratic dependence in the regime just above $\Delta=1$\footnote{While the line in this figure is not a fit we checked that vastly different power laws such as $\sim\Delta$ and $\sim\Delta^3$ are clearly inconsistent with the data.}.

While all of the above results are in agreement with those found in Ref.~[\citen{Barmettler2010}] with iMPS, the damping rates inferred do not agree quantitatively with the iMPS results. Moreover, the quantum critical point seems to be slightly shifted away from $\Delta=1$ in our approximation, as evidenced by the simple exponential damping of the $\Delta=1$ curve shown in Fig.~\ref{fig_rel_HB_NN}, instead of the oscillations around zero found in Ref.~[\citen{Barmettler2010}]. Other features not well reproduced by our approximation include the approximate $\Delta$-independence of the oscillation periods found for $\Delta<1$ and the algebraic decay expected for $\Delta=0$.

Despite these quantitative inaccuracies, which may be improved in the next order of the $1/N$ expansion, it is remarkable that our 2PI approximation is able to reproduce most generic features of the relaxation dynamics around the QPT of the XXZ chain, even in the strongly interacting regime around $\Delta=1$. In particular, it greatly outperforms previous mean-field treatments built on a mapping to spinless fermions which show a qualitatively different behavior. The results presented here open up the possibility to study dynamical quantum phase transitions in lattice spin systems in regimes in which methods such as iMPS or other DMRG related methods fail, e.g.~in higher dimensions as previously done in the $O(N)$ model~\cite{PhysRevB.96.134313}. For this purpose, our results suggest that one would not need to simulate large system sizes owing to the fast convergence to the thermodynamic limit shown here.
\section{Signatures of Many Body Localization in a Heisenberg chain\label{MBL}}

In the first two applications, we have shown that the Schwinger boson spin-2PI method is able to reproduce generic features of thermalization dynamics in interacting spin models. In this section, we give some indicative results that it is also able to capture the dynamics of local observables in a system which refuses to thermalize: a many-body localized (MBL) system \cite{BASKO20061126,Pal,PhysRevX.5.031032,PhysRevLett.113.243002,Choi1547,Smith2016}. The model best studied in this context is the Heisenberg chain with nearest-neighbour interactions in a random field~\cite{PhysRevLett.114.160401,luitz_david_j._ergodic_2017,bardarson_unbounded_2012}
\begin{equation}
\hat{H}= J \sum_{i} \left(\hat{S}_i^x\hat{S}_{i+1}^x+ \hat{S}_i^y\hat{S}_{i+1}^y+\hat{S}_i^z\hat{S}_{i+1}^z\right)+\sum_i h_i\hat{S}_i^z,
\end{equation}
where the $h_i$ are numbers drawn from a uniform random distribution in the interval $\left[-\Theta,\Theta \right]$. Note that this Hamiltonian becomes the model (\ref{eq:H_anisotropicXXZ}) studied in the previous section for $\Delta=1$ and $\Theta=0$.  As before, we consider as initial state the classical N\'{e}el state in Eq.~(\ref{eq:IC_Neel}) and study the dynamics of the staggered magnetization, Eq.~(\ref{eq:stag_magn_def}), in a system with periodic boundary conditions.
For the purpose of localization it is useful to note that for this particular initial state, the staggered magnetization can be interpreted as quantifying the correlations with the initial state by means of~\cite{Hauke2015}
\begin{equation}
\sum_i \braket{\hat{S}_i^z(t)\hat{S}_i^z(0)} = \frac{1}{2}\sum_i (-1)^i \braket{S_i^z(t)}.
\end{equation}
For thermalizing systems with initial state in the zero-magnetization sector, such as the N\'{e}el state, the correlation with the initial state, and hence the staggered magnetization, should go to zero as a relaxing system effectively forgets its initial state. In a localized system, however, memory of the initial state is retained and therefore the above quantity tends to a nonzero constant in a fully many-body localized system. 

\begin{figure}[htp]
\centering
\includegraphics{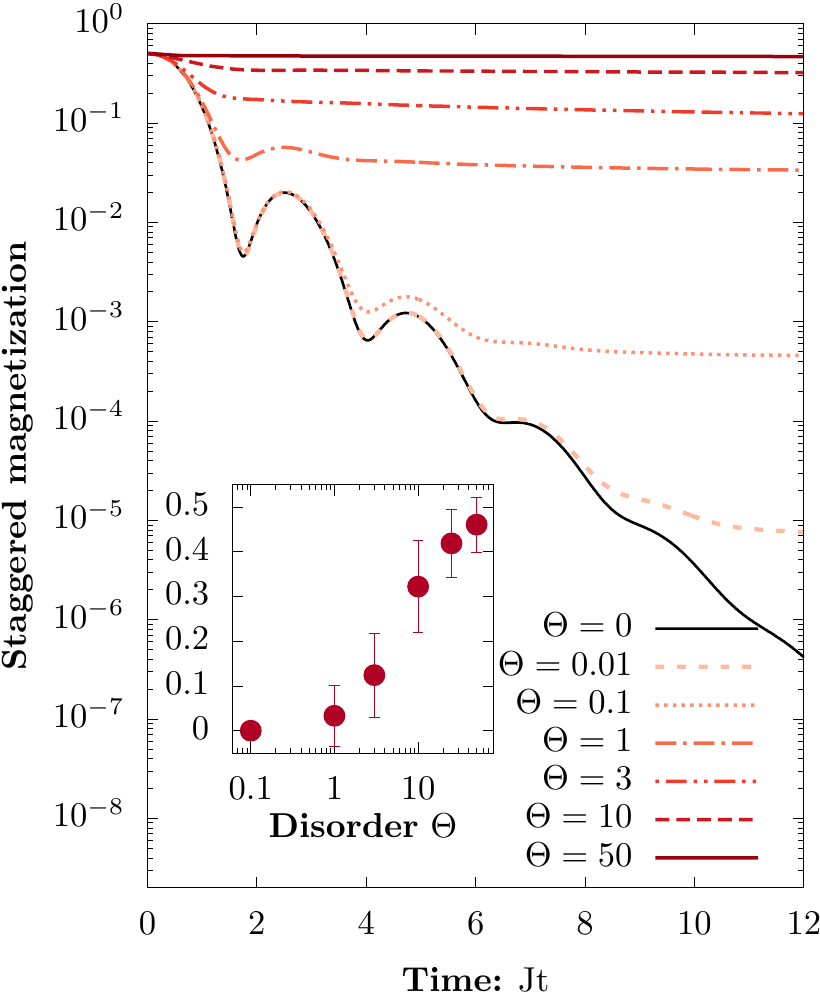}
\caption{Time evolution of the staggered magnetization in a Heisenberg chain of 6 spins with a random field for different disorder strengths $\Theta$, averaged over $26$ realizations of the disorder. As the disorder increases, the dynamics slow down from exponentially fast relaxation to a full arrest on the observed timescales. Inset: Latest value of the staggered magnetization as a function of disorder strength. Error bars are defined as the standard deviation of the disorder average.}
\label{fig_reltimes_HB_MBL}
\end{figure}
Fig.~\ref{fig_reltimes_HB_MBL} shows the time evolution of the staggered magnetization in a chain of six spins initialized in a N\'{e}el ordered state for various disorder strengths $\Theta$. The results displayed are averaged over $26$ disorder realizations. Based on the finite-size discussion of the previous section for the case without disorder (see Fig.~\ref{fig_finite_size_flow}), we expect them to capture at least some qualitative features of the system in the thermodynamic limit. For very weak disorder ($\Theta=0.01$), we observe that the time evolution is indistinguishable from the case of no disorder for early times and the relaxation slows down at around $Jt=8$. For larger disorder strengths, a plateau is approached and the value of the staggered magnetization at the plateau is found to increase with increasing disorder. For $\Theta=50$ no time evolution of the staggered magnetization is visible on the observed timescale. In the inset, we show the latest value of the staggered magnetization as a function of the disorder. A crossover from thermalization at low disorder strength to no relaxation at strong disorder is visible (see inset in Fig.~\ref{fig_reltimes_HB_MBL}), where the inflexion point resulting from interpolating between the points is consistent with the value $\Theta\approx3.5$ obtained in Ref.~[\citen{Singh2016}] for the location of the MBL transition. 

For the results shown in this section we again do {\it not} use memory cuts as in the previous section. Nevertheless, it is interesting to note that, counter-intuitively, employing a memory cut happens to work better the stronger the disorder is, even though the memory of the initial state lasts longer in this case. On a technical level, this may be understood from the fact that the disorder enters quadratically in the Schwinger Bosons (and therefore already at LO) into the action whereas interaction effects enter through the NLO self energies and therefore through the memory integrals. At weak disorder, the interactions dominate and therefore the memory integrals are important, whereas at strong disorder the opposite is the case and therefore the memory integrals can be cut. In Ref.~[\citen{MBLKnap}] this fact has been used in greater depth to develop a simple Hartree-Fock theory of the many-body-localization transition.

While these observations are in agreement with previous numerical studies of MBL in this system, we note that the observed timescales as well as the system size are not large enough to conclusively demonstrate that this method is able to describe this phenomenon.  Future studies would, however, be immediately able to generalize results to higher dimensions and more exotic interactions (such as long-range interactions), where other standard numerical methods become inapplicable. Moreover, it is useful to note that in contrast to conventional field-theoretic treatments of disordered systems~\cite{PhysRevB.60.2218}, the disorder is taken into account without further approximations as it is quadratic in the Schwinger boson operators. 


\section{Conclusions and Outlook}
Our work presents a non-equilibrium quantum field theory approach to the dynamics of arbitrary spin models using a symmetry conserving $1/N$ expansion of the 2PI effective action. Its non-perturbative nature means that our theory is not restricted to a small interaction parameter. We argue that $N$ is related to a residual $O(2)$ symmetry of our mapping of spins to Schwinger bosons and show that the Schwinger boson constraint emerges as a conserved current of this symmetry, which is not violated in 2PI approximations. We furthermore show how spin correlators can be extracted from a Hubbard Stratonovich field correlator.

We benchmark our method in various settings. First, we describe the relaxation dynamics in a 3D long-range interacting dipolar XY model with quenched disorder as implemented in Rydberg atom experiments. We find substantial improvement over the mean field solution in a system with 100 spins, a regime far from the applicability of exact diagonalization. Only small deviations from a method considered numerically exact in this regime, MACE, are found, while all qualitative features of the dynamics are recovered. Furthermore, we study the thermalization dynamics of a N\'{e}el ordered initial state on different sides of a (dynamical) quantum phase transition in a 1D (an)isotropic XXZ model in a regime in which the mean-field approximation does not show any dynamics. We find that our method reproduces most qualitative features found previously with matrix product states. Lastly, we give some indicative results that our method is able to describe the transition from a thermalizing to a many-body-localized phase in a 1D Heisenberg chain in a random field.

These benchmarks show that our non-equilibrium quantum field theory method is able to describe generic features found in the local magnetization dynamics of strongly correlated spin models implemented in current cold atom experiments, such as models with quenched  disorder in interactions and/or external fields, long-range interactions and quantum phase transitions. Furthermore, it is not restricted to small system sizes or low dimensionality and its quickly converging finite size flow leads to the capability of extracting the time evolution in the thermodynamic limit. Our description in terms of Schwinger bosons could furthermore be used to study the quantum-classical crossover by studying the dependence of the dynamics on the spin length.

This opens up a whole range of possible applications, most notably to the thermalization dynamics of local observables in systems exhibiting a many-body-localization transition as well as dynamical quantum phase transitions. Furthermore, the influence of dimensionality and long-range interactions on these phenomena could be examined. As the external magnetic field could in principle be made time dependent, also the order parameter dynamics in periodically driven (Floquet) systems could be examined as previously done with 2PI methods in the $O(N)$ model~\cite{Weidinger2017}. Moreover, our method can provide (at least qualitative) predictions for quantum simulation experiments for example with Rydberg atoms in optical tweezers, cold atoms in quantum gas microscopes, trapped ions or NV centers in diamond in regimes in which other methods are not available.

Our method can be extended in several ways. As an extension of the $1/N$ expansion to NNLO is numerically very expensive~\cite{PhysRevD.74.025004,PhysRevD.78.125028}, a better approximation of the initial state in terms of a non-Gaussian state could have more potential for substantial improvement. Furthermore, the use of more efficient numerical algorithms might enable the evaluation of the inhomogeneous 2PI equations for system sizes close to the thermodynamic limit also in 3D.

\begin{acknowledgments}
We thankfully acknowledge discussions with Ignacio Aliaga Sirvent, Eleanor Crane, Oscar Garcia-Montero, Philipp Hauke, Michael Knap, Alexander Rothkopf, Simon Weidinger and Torsten Zache. A.S. acknowledges financial support from the International Max Planck Research School for Quantum Science and Technology (IMPRS-QST). This work is part of and supported by the DFG Collaborative Research Centre ``SFB 1225 (ISOQUANT)''. Parts of this work were performed on the computational resource bwUniCluster funded by the Ministry of Science, Research and the Arts Baden-W\"urttemberg and the Universities of the State of Baden-W\"urttemberg, Germany, within the framework program bwHPC. The authors gratefully acknowledge the compute and data resources provided by the Leibniz Supercomputing Centre (www.lrz.de).
\end{acknowledgments}

\appendix

\section{\texorpdfstring{$\mathcal{K}$}{K} identity\label{app:Kidentity}}
We prove the following identity between $\mathcal{K}$ and the totally antisymmetric tensor $\epsilon$:
\begin{equation}
\mathcal{K}_{ac}^{\alpha}\mathcal{K}_{bd}^{\beta}\left[\mathbb{1}\otimes\sigma_y\right]^{dc}=i\sum_\gamma \epsilon^{\alpha\beta\gamma}\mathcal{K}_{ab}^{\gamma}.
\label{eq:Kidentity}
\end{equation}
The proof is based on the comparison of the spin commutation relations written with Schwinger bosons and with spin variables. Firstly,
\begin{align}
\left[\hat{S}^{\alpha}_i,\hat{S}^{\beta}_j\right]=&\,i\delta_{ij}\sum_\gamma \epsilon^{\alpha\beta\gamma}\hat{S}^{\gamma}_i \nonumber\\
	=&\, i\delta_{ij}\sum_{\gamma}\epsilon^{\alpha\beta\gamma}\frac{1}{4}\mathcal{K}_{ab}^{\gamma}\hat{\varphi}_i^a\hat{\varphi}_j^b,
\end{align}
where we have inserted the real Schwinger boson representation~(\ref{eq:SB_trafo_realphi}) after using the spin commutation relations. We can however also perform these steps in reverse order, giving
\begin{align*}
\left[\hat{S}^{\alpha}_i,\hat{S}^{\beta}_j\right]&=\frac{1}{16}\mathcal{K}_{ab}^{\alpha}\mathcal{K}_{cd}^{\beta}\left[\hat{\varphi}_i^a\hat{\varphi}_i^b,\hat{\varphi}_j^c\hat{\varphi}_j^d\right] \\
&= \frac{1}{4}\mathcal{K}_{ab}^{\alpha}\mathcal{K}_{cd}^{\beta}\delta_{ij}\hat{\varphi}_i^a\hat{\varphi}_i^c\left[\mathbb{1}\otimes\sigma_y\right]^{db}, \numberthis
\end{align*}
where in the last step the commutation relations for the Schwinger boson field~(\ref{eq:4_SB_real_comm}) and the symmetry of the $\mathcal{K}_{ab}^{\alpha}$ was used.
Comparing both of the above expressions for the commutator leads to the identity (\ref{eq:Kidentity}).

\section{Ward-Takahashi identities\label{app:WTI}}
In this section, we derive a set of Ward-Takahashi identities (WTI) for the current associated to the $U(1)$ symmetry of the complex Schwinger boson action~(\ref{eq:action_complex}) in complete analogy to the textbook derivation of Ward identities in, e.g., QED~\cite{Peskin95}. As we will see, the identities $\partial_t \braket{(\hat n_i(t))^k}=0$, $\forall k\in\mathbb{N}$, which follow from the Schwinger boson constraint~(\ref{eq:SB_constraint}), correspond to special cases of these WTIs.

We start by transforming the Schwinger bosons according to the following infinitesimal transformation
\begin{equation}
\psi^a_i(t)\rightarrow \psi'^a_i(t)=\psi^a_i(t)+i\alpha_i(t)\psi^a_i(t),
\end{equation}
where we work in the complex basis for convenience.
Note that we make $\alpha_i(t)$ explicitly time dependent.
The measure of the functional integration  is invariant under such a unitary transformation such that this transformation merely acts like a change of coordinates and it follows that
\begin{equation}
\int\mathcal{D}[\bar\psi,\psi] \exp{\lbrace iS[\bar\psi,\psi]\rbrace} =\int\mathcal{D}[\bar\psi,\psi] \exp{\lbrace iS[\bar\psi',\psi']\rbrace}.
\end{equation}
Expanding the RHS to first order in $\alpha$ leads to
\begin{align}
0 &= \int\mathcal{D}[\bar\psi,\psi] \exp{\lbrace iS[\bar\psi,\psi]\rbrace}\times\nonumber\\&\qquad\times\left(-i\sum_j \int_\mathcal{C} \mathrm{d}t \left(\partial_t\alpha_j(t) \bar\psi_j^a \psi_j^a\right)\right) \nonumber\\
&= \int\mathcal{D}[\bar\psi,\psi] \exp{\lbrace iS[\bar\psi,\psi]\rbrace}\left(-i\sum_j \int_\mathcal{C} \mathrm{d}t\,\alpha_j(t) \partial_tj_j^0\right) ,
\end{align}
where it was used that the variation with respect to $\alpha$ vanishes as $S$ is invariant under transformations with constant $\alpha$. The only non-vanishing contribution is therefore the variation with respect to $\partial_t\alpha$ of the kinetic term. In the second step, partial integration was used and the classical Noether current $j^0_i = - n_i\equiv - \bar\psi_i^a \psi_i^a$ was inserted.
Noting that the expression must hold for arbitrary $\alpha_j(t)$ and dividing by $Z=\int\mathcal{D}[\bar\psi,\psi] \exp{\lbrace iS\rbrace}$ we can follow
\begin{equation}
\label{firstWTI}
\partial_t\braket{j_j^0(t)} = 0 \Rightarrow \partial_t \braket{\hat{n}_i(t)} =0,
\end{equation}
i.e.~the expecation value of the Schwinger boson number operator is a constant, also in the exact quantum theory.

The same can now be done for the expectation value with two field insertions, i.e.
\begin{align}
&\int\mathcal{D}[\bar\psi,\psi] \exp{\left\lbrace iS[\bar\psi,\psi]\right\rbrace} \psi^a_i(t_1)\bar\psi^b_j(t_2) \nonumber\\
&=\int\mathcal{D}[\bar\psi,\psi] \exp{\left\lbrace iS[\bar\psi',\psi']\right\rbrace}\psi'^a_i(t_1)\bar\psi'^b_j(t_2).
\end{align}
Again expanding to first order in $\alpha$ leads to
\begin{align}
0 &= \int\mathcal{D}[\bar\psi,\psi] \exp{\left\lbrace iS[\bar\psi,\psi]\right\rbrace} \psi^a_i(t_1)\bar\psi^b_j(t_2)\times\nonumber\\
&\quad \times\left(-i\sum_k \int_\mathcal{C} \mathrm{d}t \alpha_k(t) \partial_tj_k^0(t)+i\alpha_i(t_1)-i\alpha_j(t_2)\right).
\end{align}
By introducing (contour-) delta functions we can extend the sum over $k$ and the contour time integral over the whole bracket. Argueing again as above, this results in the second WTI
\begin{align}
\label{2_WTI2}
&\braket{\partial_tj^0_k(t)\psi^a_i(t_1)\bar\psi^b_j(t_2) } \nonumber\\
&=\braket{ \psi^a_i(t_1)\bar\psi^b_j(t_2)(\delta_{ik}\delta_\mathcal{C}(t-t_1)-\delta_{jk}\delta_\mathcal{C}(t-t_2))}.
\end{align}
To relate this back to the quantity $\hat n_i$, we consider the special case $i=j$, $t_1=t_2\equiv t$, for which the second WTI becomes
\begin{equation}
	\braket{(\partial_t j^0_k(t))\psi^a_i(t)\bar\psi^b_i(t) } = 0 .
\label{eq:2_WTI2_special}
\end{equation}
Suppressing time arguments, the time derivative of the constraint squared can then be re-expressed as
\begin{align}
\partial_t \braket{\hat{n}^2_i(t)} &=\partial_t\braket{\hat\psi_i^{a\dagger} \hat\psi_i^a\hat\psi_i^{b\dagger} \hat\psi_i^b} \nonumber\\
&= \partial_t\braket{\bar\psi_i^{a} \psi_i^a \bar\psi_i^{b} \psi_i^b}\nonumber\\
&=2\braket{\bar\psi_i^{a}\psi_i^a\partial_t(\bar\psi_i^{b}\psi_i^b)}\nonumber\\
&= 0,
\label{eq:constr2_to_WTI}
\end{align}
where in the second line the operator expectation value was rewritten in terms of a path-integral expectation value
\footnote{We note that the equal-time expectation value $\braket{\bar\psi_i^a(t) \psi_j^b(t)}$ corresponds to the symmetrically ordered product of operators $\frac{1}{2} \braket{ \{ \hat\psi_i^{a\dagger}(t), \hat\psi_j^{b}(t) \}}$. Similarly, higher-order products of fields taken at equal times correspond to symmetrically ordered products of operators. This has to be taken into account when writing the expectation value $\braket{(\hat n_i)^k}$ in terms of path-integral expectation values of the fields $\bar\psi$, $\psi$. In particular, this affects the derivation of $\partial_t \braket{(\hat n_i(t))^k}=0$ from the $k$-th order WTI presented here. However, it can straightforwardly be shown that the derivation remains valid as long as one can write $\braket{(\hat n_i)^k}$ as a linear combination of $\braket{(n_i)^q} \equiv \braket{ (\bar\psi_i^{a} \psi_i^a )^q}$ with $q\leq k$. We explicitly checked that this is indeed the case up to $k=4$.},
and we made use of the first WTI, $\partial_t\braket{\hat n_i(t)}=0$.

Higher-order WTIs may be obtained analogously. When considering the special case of insertions of the form $(\psi^a_i(t)\bar\psi^b_i(t) )^{k-1}$, the resulting WTI can be related, as in Eq.~(\ref{eq:constr2_to_WTI}), to the identity $\partial_t \braket{(\hat n_i(t))^k}=0$\cite{Note2}.

\section{NLO: Homogeneous initial states}
\label{Init_state_hom}
For spatially homogeneous systems with periodic boundary conditions, i.e.~translationally invariant initial states and interactions, $G$, $\Sigma$ and $\Pi$ become independent of the lattice site. The correlator $D$ and the interaction matrix $J$ depend only on the distance between two sites and can be Fourier transformed with momentum $\mbf k$ as
\begin{equation}
	J_{\mbf k} = \sum_{\mbf x} e^{i\mbf k \mbf x} J_{\mbf x},
\end{equation}
and similarly for $D$. The inverse transform is normalized by $1/N_s$, where $N_s$ is the number of spins. Using this, the equations of motion for the correlators, Eqs.~(\ref{eq:eomF_NLO}), (\ref{eq:eomrho_NLO}), can then be simplified to
\begin{align}
\partial_{t_1}F^{ab}(t_1,t_2) =&\, i\left[\mathbb{1}\otimes\sigma_y\right]^{ac}\bigg\lbrace\frac{1}{2} \sum_{\gamma}\bar{\chi}^{\gamma}(t_1)\mathcal{K}_{cd}^{\gamma}F^{db}(t_1,t_2)\nonumber\\
+&\int_0^{t_1}\mathrm{d}t\,\Sigma^{\rho,cd}(t_1,t)F^{db}(t,t_2)\nonumber\\
-&\int_0^{t_2}\mathrm{d}t\,\Sigma^{F,cd}(t_1,t)\rho^{db}(t,t_2)\bigg\rbrace,\label{eq_F_hom}\\
\partial_{t_1}\rho^{ab}(t_1,t_2)=&\, i\left[\mathbb{1}\otimes\sigma_y\right]^{ac}\bigg\lbrace\frac{1}{2}\sum_{\gamma}\bar{\chi}^{\gamma}(t_1)\mathcal{K}_{cd}^{\gamma}\rho^{db}(t_1,t_2)\nonumber\\
+&\int_{t_2}^{t_1}\mathrm{d}t\, \Sigma^{\rho,cd}(t_1,t)\rho^{db}(t,t_2)\bigg\rbrace,\label{eq_Rho_hom}
\end{align}
with Schwinger boson self energies, Eqs.~(\ref{4_Sigma_F}), (\ref{4_Sigma_rho}), given by
\begin{align}
&\Sigma^{F,ab}(t_1,t_2)= -\frac{1}{4N_s}\sum_{\alpha\beta}\mathcal{K}_{ac}^{\alpha}\mathcal{K}_{bd}^{\beta}\sum_\mathbf{k}J_{\mathbf{k}}^{\alpha}J_{\mathbf{k}}^{\beta}\nonumber\\&\times\bigg(F^{cd}(t_1,t_2)\hat{D}_{\mathbf{k}}^{F,\alpha\beta}(t_1,t_2)-\frac{1}{4}\rho^{cd}(t_1,t_2)\hat{D}^{\rho,\alpha\beta}_{\mathbf{k}}(t_1,t_2)\bigg), \\ 
&\Sigma^{\rho,ab}(t_1,t_2)= -\frac{1}{4N_s}\sum_{\alpha\beta}\mathcal{K}_{ac}^{\alpha}\mathcal{K}_{bd}^{\beta}\sum_\mathbf{k}J_{\mathbf{k}}^{\alpha}J_{\mathbf{k}}^{\beta}\nonumber\\&\times\bigg(\rho^{cd}(t_1,t_2)\hat{D}_{\mathbf{k}}^{F,\alpha\beta}(t_1,t_2)+F^{cd}(t_1,t_2)\hat{D}^{\rho,\alpha\beta}_{\mathbf{k}}(t_1,t_2)\bigg).
\end{align}
The equations of motion for the auxiliary field, Eqs.~(\ref{eq:eomDF_NLO}), (\ref{eq:eomDrho_NLO}), simplify to
\begin{align*}
&\hat{D}^{F,\alpha\beta}_{\mathbf{k}}(t_1,t_2) = -\Pi^{F,\alpha\beta}(t_1,t_2)\nonumber\\&\qquad+\int_0^{t_1}\mathrm{d}t \,\sum_{\delta}\Pi^{\rho, \alpha\delta}(t_1,t)J_{\mathbf{k}}^{\delta}\hat{D}^{F,\delta\beta}_{\mathbf{k}}(t,t_2)\nonumber\\&\qquad-\int_0^{t_2} \mathrm{d}t\,\sum_{\delta}\Pi^{F,\alpha\delta}(t_1,t)J_{\mathbf{k}}^{\delta}\hat{D}^{\rho,\delta\beta}_{\mathbf{k}}(t,t_2), \numberthis \\
&\hat{D}^{\rho,\alpha\beta}_{\mathbf{k}}(t_1,t_2) = -\Pi^{\rho,\alpha\beta}(t_1,t_2)\nonumber\\&\qquad+\int_{t_2}^{t_1}\mathrm{d}t\, \sum_{\delta}\Pi^{\rho,\alpha\delta}(t_1,t)J_{\mathbf{k}}^{\delta}\hat{D}^{\rho,\delta\beta}_{\mathbf{k}}(t,t_2), \numberthis
\end{align*}
with auxiliary field self energies, Eqs.~(\ref{4_Pi_F}), (\ref{4_Pi_rho}), resulting as
\begin{align}
&\Pi^{F,\alpha\beta}(t_1,t_2) = -\frac{1}{8}\mathcal{K}_{ab}^{\alpha}\mathcal{K}_{cd}^{\beta}\bigg(F^{ac}(t_1,t_2)F^{bd}(t_1,t_2)\nonumber\\&\qquad-\frac{1}{4}\rho^{ac}(t_1,t_2)\rho^{bd}(t_1,t_2)\bigg), \\
&\Pi^{\rho,\alpha\beta}(t_1,t_2) = -\frac{1}{4}\mathcal{K}_{ab}^{\alpha}\mathcal{K}_{cd}^{\beta}F^{ac}(t_1,t_2)\rho^{bd}(t_1,t_2),
\end{align}
and the auxiliary field one-point function, Eq.~(\ref{eq:eomChi_NLO}), as
\begin{equation}
\bar{\chi}^{\alpha}(t)=\frac{1}{4} J_{\mbf k=0}^{\alpha}\mathcal{K}^{\alpha}_{cd}F^{cd}(t,t).
\end{equation}
In the above equations we omitted possible external fields which can be incorporated by replacing
\begin{equation}
\bar{\chi}^{\gamma}(t_1)\rightarrow\bar{\chi}^{\gamma}(t_1)+B^{\gamma}
\end{equation}
in Eqs.~(\ref{eq_F_hom}), (\ref{eq_Rho_hom}).

\section{Spin Observables from Auxiliary Field Correlators\label{Chapter_SpinObsfromAux}}
In this appendix we clarify the connection between the auxiliary field $\chi$ and spin variables. Loosely speaking, we aim to establish the following links:
\begin{equation}
\braket{\chi} \leftrightarrow \braket{S}, \qquad\braket{\chi\chi} \leftrightarrow \braket{SS}.
\end{equation}
For this purpose, we start from the auxiliary field action in Eq.~(\ref{eq:action_chi}) and introduce a source field $\eta_j^{\alpha}$ for the auxiliary field via
\begin{equation}
S[\varphi,\chi]\rightarrow S[\varphi,\chi]+i\int_{\mathcal{C}}\mathrm{d}t \sum_{j,\alpha} \eta_j^{\alpha}\chi_j^{\alpha}.
\end{equation}
In order to understand the relation between functional derivatives with respect to this source field and spin expectation values, we first complete the squares in the interaction part of the action as follows:
\begin{widetext}
\begin{align*}
S_{\mathrm{int}}[\varphi,\chi] &= \int_\mathcal{C}\mathrm{d}t \left\lbrace\frac{1}{2}\sum_{jk,\alpha}\left[J^{-1}\right]_{jk}^{\alpha}\left(\chi_j^{\alpha}-\sum_m J_{jm}^{\alpha}\left(\frac{1}{4}\mathcal{K}_{ab}^{\alpha}\varphi^a_m\varphi_m^b-\eta_m^{\alpha}\right)\right)\left(\chi_k^{\alpha}-\sum_l J_{kl}^{\alpha}\left(\frac{1}{4}\mathcal{K}_{cd}^{\alpha}\varphi^c_l\varphi_l^d-\eta_l^{\alpha}\right)\right)\right\rbrace \\
&\quad+\int_\mathcal{C}\mathrm{d}t \left\lbrace-\frac{1}{2}\sum_{ij,\alpha}J_{ij}^{\alpha}\left(\frac{1}{4}\mathcal{K}_{ab}^{\alpha}\varphi^a_i\varphi_i^b-\eta_i^{\alpha}\right)\left(\frac{1}{4}\mathcal{K}_{cd}^{\alpha}\varphi^c_j\varphi_j^d-\eta_j^{\alpha}\right)\right\rbrace.\numberthis
\end{align*}
\end{widetext}
Note that we assume $B=0$ in the following, but the results can be straightforwardly generalized to nonzero external field. Shifting the source field by $\chi_j^{\alpha}\rightarrow \chi_j^{\alpha}+\sum_m J_{jm}^{\alpha}\left(\frac{1}{4}\mathcal{K}_{ab}^{\alpha}\varphi^a_m\varphi_m^b-\eta_m^{\alpha}\right)$ we can integrate out the auxiliary field by standard Gaussian functional integration, which yields
\begin{align}
Z[\eta] &= \int\mathcal{D}\varphi\int\mathcal{D}\chi \exp\left(i S[\varphi,\chi]+i\int_{\mathcal{C}}\mathrm{d}t \sum_{j,\alpha} \eta_j^{\alpha}\chi_j^{\alpha}\right)\notag\\
&\propto \int\mathcal{D}\varphi \exp\bigg(iS[\varphi]+i\int_\mathcal{C}\mathrm{d}t \sum_{ij,\alpha}J_{ij}^{\alpha}\nonumber\\&\qquad\qquad\times\Big(\eta_i^{\alpha}\frac{1}{4}\mathcal{K}_{cd}^{\alpha}\varphi^c_j\varphi_j^d-\frac{1}{2}\eta_i^{\alpha}\eta_j^{\alpha}\Big)\bigg),
\end{align}
where the proportionality factor is the determinant from the Gaussian integral. $S[\varphi]$ denotes the Schwinger boson action before introducing the auxiliary field, which is given in Eq.~(\ref{eq:action_reals}).

In the next two subsections, we derive the relationship between the one- and two-point functions of the auxiliary field and spin expectation values. The strategy consists in writing functional derivatives of $Z[\eta]$ with respect to $\eta$, which define expectation values of $\chi$, in terms of $\varphi$ correlators. The latter can then be associated to operator expectation values of $\hat\varphi$, which are in turn related to spin variables $\hat S$ by the Schwinger boson mapping (\ref{eq:SB_trafo_realphi}).
In the following, brackets of field variables will refer to averages with respect to the path integral with $S[\varphi]$, i.e.
\begin{equation}
\braket{(\cdot)} = \int\mathcal{D}\varphi (\cdot) \exp(iS[\varphi]),
\end{equation}
where the ordering along the closed time contour needs to be taken into account.

We note that the relations derived here are strictly only valid in the exact theory, and deviations can be expected when employing approximations to the effective action. In appendix \ref{ChecksAuxCorr} we therefore check whether standard relations between spin correlators are reproduced by the corresponding $\chi$ correlators at NLO, and argue how deviations from the expected results might be overcome by future work.
\subsection{One-Point Function}
The one point function of the auxiliary field can be obtained by deriving the generating functional once with respect to the source field, i.e.
\begin{align}
\braket{\chi_i^{\alpha}(t)}&=\frac{1}{Z}\frac{\delta Z[\eta]}{i\delta \eta_i^{\alpha}(t)}\bigg|_{\eta=0}\\ &= \sum_k J_{ik}^{\alpha} \bigg\langle\frac{1}{4}\mathcal{K}_{ab}^{\alpha}\varphi^a_k\varphi_k^b\bigg\rangle\\ &= \sum_k J_{ik}^{\alpha} \braket{\hat{S}^{\alpha}_k},
\end{align}
which just reproduces the result obtained from the equation of motion for the auxiliary field, see Eq.~(\ref{Chi_eom}). Multiplying from the left with the inverse interaction matrix (assuming it is invertible) gives the sought expression for the spin one-point-function,
\begin{equation}
\label{Chi_spin_mag}
\braket{\hat{S}^{\alpha}_i} = \sum_k \left[J^{-1}\right]_{ik}^{\alpha} \braket{\chi_k^{\alpha}(t)} .
\end{equation}
We have tested this analytical identity in our numerical evaluations by comparing the result for the magnetizations obtained from the auxiliary field (\ref{Chi_spin_mag}) with the one from the Schwinger boson two-point function (\ref{mag_S}) and found agreement between the two. 

\subsection{Two-Point Function}
Similarly, one can calculate the two-point-function by deriving the generating functional twice, 
\begin{align}
&\braket{\chi_i^{\alpha}(t_1)\chi_j^{\beta}(t_2)} = \frac{1}{Z}\frac{\delta Z[\eta]}{i\delta \eta_i^{\alpha}(t_1)i\delta \eta_j^{\beta}(t_2)}\bigg|_{\eta=0}\notag\\
&= i J_{ij}^{\alpha}\delta^{\alpha\beta}\delta_\mathcal{C}(t_1-t_2)+\sum_{ml}J^{\alpha}_{im}J^{\beta}_{jl}\times\notag\\
&\qquad\times\bigg\langle\frac{1}{4}\mathcal{K}_{ab}^{\alpha}\varphi^a_m(t_1)\varphi_m^b(t_1)\frac{1}{4}\mathcal{K}_{cd}^{\beta}\varphi^c_l(t_2)\varphi_l^d(t_2)\bigg\rangle \notag\\
&= i J_{ij}^{\alpha}\delta^{\alpha\beta}\delta_\mathcal{C}(t_1-t_2) + \sum_{ml}J^{\alpha}_{im}J^{\beta}_{jl}\braket{T_\mathcal{C}\hat{S}_m^{\alpha}(t_1)\hat{S}_l^{\beta}(t_2)}. \label{Eq_RHS}
\end{align}
Using the definition (\ref{4_def_D}) and the decomposition of Eqs.~(\ref{4_D_decomp}), (\ref{decomp_DF_Drho}), the left-hand side of Eq.~(\ref{Eq_RHS}) can be written as
\begin{align}
\mathrm{LHS} &= D^{\alpha\beta}_{ij}(t_1,t_2) + \bar{\chi}_i^{\alpha}(t_1)\bar{\chi}_j^{\beta}(t_2)\notag\\
&= i J_{ij}^{\alpha}\delta^{\alpha\beta}\delta_\mathcal{C}(t_1-t_2)+ \sum_{kl} J_{ik}^{\alpha}J_{lj}^{\beta}\notag\\
&\quad\times\bigg\lbrace\hat{D}_{kl}^{F,\alpha\beta}(t_1,t_2)-\frac{i}{2}\sgn_\mathcal{C} (t_1-t_2)\hat{D}_{kl}^{\rho,\alpha\beta}(t_1,t_2)\notag\\
&\qquad\quad+\braket{\hat{S}^{\alpha}_k(t_1)}\braket{\hat{S}^{\beta}_l(t_2)}\bigg\rbrace .
\end{align}
Similarly, we decompose the time ordered spin correlator [c.f.~Eq.~(\ref{decomp_F_Rho})] into anticommutator and commutator parts as $\braket{T_\mathcal{C}\hat{S}\hat{S}}_\mathrm{C} = F^S-\frac{i}{2}\sgn_\mathcal{C}\rho^S$ with
\begin{align}
	F^{S,\alpha\beta}_{ij} (t_1,t_2) \equiv&\, \frac{1}{2}\braket{\left\lbrace \hat S_i^\alpha(t_1), \hat S_j^\beta(t_2)\right\rbrace}_c, \\
	\rho^{S,\alpha\beta}_{ij}(t_1,t_2) \equiv&\, i\braket{\left[ \hat S_i^\alpha(t_1), \hat S_j^\beta(t_2)\right]}.
\end{align}
In this way, the right-hand side of (\ref{Eq_RHS}) becomes
\begin{align}
&\mathrm{RHS} = i J_{ij}^{\alpha}\delta^{\alpha\beta}\delta_\mathcal{C}(t_1-t_2) + \sum_{kl} J_{ik}^{\alpha} J_{jl}^{\beta}\times\notag\\
&\times\bigg\lbrace
\left(F^{S,\alpha\beta}_{kl}(t_1,t_2)-\frac{i}{2}\sgn_\mathcal{C}(t_1-t_2) \rho^{S,\alpha\beta}_{kl}(t_1,t_2)\right)\notag\\
&\qquad\quad+ \braket{\hat{S}^{\alpha}_k(t_1)}\braket{\hat{S}^{\beta}_l(t_2)}\bigg\rbrace .
\end{align}
Comparing the terms on the LHS and RHS, we can therefore conclude that
\begin{align}
F^{S,\alpha\beta}_{ij}(t_1,t_2) &= \hat{D}_{ij}^{F,\alpha\beta}(t_1,t_2)
\label{F_ss},\\
\rho^{S,\alpha\beta}_{ij}(t_1,t_2) &= \hat{D}_{ij}^{\rho,\alpha\beta}(t_1,t_2),
\label{Rho_ss}
\end{align}
as given in Eqs.~(\ref{eq:spin2ptF_phichi}) and (\ref{eq:spin2ptrho_phichi}).
Note that only those components $\beta$ of the spin-spin correlators for which $J^\beta$ is invertible can be read out with Eqs.~(\ref{F_ss}), (\ref{Rho_ss}). For instance, if there is no $\hat S^z\hat S^z$ term in the Hamiltonian, the $F^{S,zz}$ component can not be obtained from the above equations. In such cases, spin-spin correlators can be computed from a Bethe-Salpeter equation approach, as described in Ref.~[\citen{Babadi15}]. We note again that both approaches are equivalent in the exact theory, but differences may arise  when doing approximations.

\subsection{Spin identities from auxiliary field correlators at NLO\label{ChecksAuxCorr}}

The relations between auxiliary field and spin correlators derived in section \ref{Chapter_SpinObsfromAux} only hold in the exact theory. In this section, we investigate whether standard relations between correlation functions imposed by the properties of the spin operators are reproduced by the corresponding $\chi$ correlators at NLO in the $1/N$ approximation. In order to distinguish the two, we denote the latter with a tilde, e.g. $\widetilde{\rho}^S$ is the spin commutator expectation value in Eq.~(\ref{Rho_ss}) as obtained from the approximated $\hat D^\rho$ to NLO. Note that a further check consists in comparing the expression for the total energy $\langle \hat{H} \rangle$ of the system in terms of spin expectation values to the corresponding expression for the energy expressed with auxiliary field correlators as obtained from the $1/N$ approximation. This is discussed in App.~\ref{app:energy}.

\subsection*{Spin commutation relations}

First, we consider the spin equal-time commutation relations, which require that the spin commutator obtained from (\ref{Rho_ss}) fulfils
\begin{equation}
\rho^{S,\alpha\beta}_{ij}(t,t)=i\braket{\left[\hat{S}_i^{\alpha}(t),\hat{S}^{\beta}_j(t)\right]} = - \delta_{ij} \sum_\gamma \epsilon^{\alpha\beta\gamma}\braket{\hat{S}^{\gamma}_i(t)}.
\end{equation}
To check this, we start from (\ref{Rho_ss}) and use the (exact) equation of motion for $D^{\rho}$, Eq.~(\ref{eq:eomDrho_NLO}), to obtain
\begin{align*}
\widetilde{\rho}^{S,\alpha\beta}_{ij}(t,t) &= -\Pi_{ii}^{\rho,\alpha\beta}(t,t) \delta_{ij}, \numberthis
\end{align*}
where the memory integral vanishes at equal-times.
Next, we insert the auxiliary field self energy in the NLO approximation, Eq.~(\ref{4_Pi_rho}), and get
\begin{align*}
\widetilde{\rho}^{S,\alpha\beta}_{ij}(t,t) &= \frac{1}{4}\mathcal{K}_{ab}^{\alpha}\mathcal{K}_{cd}^{\beta}F^{ac}_{ii}(t,t)\rho^{bd}_{ii}(t,t)\delta_{ij} \\
&= \frac{1}{4}\mathcal{K}_{ab}^{\alpha}\mathcal{K}_{cd}^{\beta}F^{ac}_{ii}(t,t)\left(-i\right)\left[\mathbb{1}\otimes\sigma_y\right]^{bd}\delta_{ij}\\
&= -\frac{1}{4} \sum_\gamma \mathcal{K}^{\gamma}_{ac}F^{ac}_{ii}(t,t)\epsilon^{\alpha\beta\gamma}\delta_{ij},\numberthis
\end{align*}
where we used the commutation relations of the Schwinger bosons, Eq.~(\ref{eq:IC_rho}), and in the last step we employed the identity (\ref{S_matrix_identity}) proved in App.~\ref{app:Kidentity}.
Inserting (\ref{mag_S}), we finally arrive at the sought identity
\begin{equation}
\widetilde{\rho}^{S,\alpha\beta}_{ij}(t,t) = - \delta_{ij}\epsilon^{\alpha\beta\gamma}\braket{\hat{S}^{\gamma}_i(t)} =\rho^{S,\alpha\beta}_{ij}(t,t). \numberthis
\end{equation}
The validity of this identity already at NLO in the $1/N$ expansion can be understood from the fact that higher orders in $1/N$ lead to memory integral terms in the self-energy $\Pi^\rho$, which vanish at equal times and hence yield no further contribution to $\widetilde\rho^S(t,t)$.

\subsection*{Initial state correlations}

Since at the initial time also the memory integrals in $\hat D^F$ vanish, we can calculate the spin spin correlations of the intitial state analytically as given by
\begin{equation}
F^{S,\alpha\beta}_{ij}(0,0) = \braket{\hat{S}^{\alpha}_i(0)\hat{S}^{\beta}_j(0)}_\mathrm{C} = 0 \quad \mathrm{for}  \quad i \neq j.
\end{equation}
Using the equations of motion for $\hat D^F$, Eq.~(\ref{eq:eomDF_NLO}), we obtain
\begin{align}
\widetilde{F}^{S,\alpha\beta}_{ij}(0,0)
&= -\Pi_{ii}^{F,\alpha\beta}(0,0) \,\delta_{ij}\label{F_S_Pi_F} \nonumber\\
&= 0 \qquad \mathrm{for} \qquad i \neq j.
\end{align}
The vanishing of the connected correlators implies that the initial state in the Gaussian approximation is a product state as expected.

\subsection*{Spin length constraint at initial time}
Another condition which needs to be fulfilled is the spin length constraint, i.e.
\begin{equation}
\braket{\hat{\vec{S}}^2_i(t)} = S(S+1).
\end{equation}
At the initial time, $t=0$, we can check this analytically, i.e.~we will check whether
\begin{equation}
\sum_{\alpha} \left(F^{S,\alpha\alpha}_{ii}(0,0) + \braket{\hat{S}^{\alpha}_i(0)}^2 \right)= S(S+1)
\end{equation}
is also true for $F^{S}\rightarrow\widetilde{F}^{S}$. We use Eq.~(\ref{F_S_Pi_F}) to relate the spin-spin correlator to the auxiliary field self-energy $\Pi$ and insert the corresponding expression at NLO, Eq.~(\ref{4_Pi_F}), so that
\begin{align}
&\widetilde{F}^{S,\alpha\alpha}_{ii}(0,0) \nonumber\\&= \frac{1}{8}\mathcal{K}_{ab}^{\alpha}\mathcal{K}_{cd}^{\alpha}\left(F^{ac}_{ii}(0,0)F^{bd}_{ii}(0,0)-\frac{1}{4}\rho^{ac}_{ii}(0,0)\rho^{bd}_{ii}(0,0)\right) \nonumber\\
&= \frac{1}{8}\bigg(\Tr{\left[\mathcal{K}^{\alpha}F_{ii}(0,0)\mathcal{K}^{\alpha}F_{ii}(0,0)\right]}\nonumber\\&\qquad\qquad+\frac{1}{4}\Tr{\left[\mathcal{K}^{\alpha}\rho_{ii}(0,0)\mathcal{K}^{\alpha}\rho_{ii}(0,0)\right]}\bigg),
\end{align} 
where in the second equality we have used the symmetry properties of $\mathcal{K}$, $F$, and $\rho$ and the trace runs over the Schwinger boson indices.

Inserting the expressions for the initial time $F$ and $\rho$, Eqs.~(\ref{eq:IC_F}), (\ref{eq:IC_rho}), the explicit form of $\mathcal{K}^{\alpha}$, Eq.~(\ref{eq:S_matrix_def}), and performing the traces then leads to
\begin{align}
\sum_{\alpha}\braket{\widetilde{S}_i^{\alpha}(0)^2} &\equiv\sum_{\alpha} \left(\widetilde F^{S,\alpha\alpha}_{ii}(0,0) + \braket{\hat{S}^{\alpha}_i(0)}^2 \right)\notag\\&= 1.5\,S(S+1) - \frac{1}{2}\sum_{\alpha}\braket{\hat{S}_i^{\alpha}(0)}^2 \nonumber\\
&\stackrel{\text{i.g.}}{\neq} S(S+1),
\label{eq:lengthconstr_broken}
\end{align}
which shows that the length constraint is, in general, not exactly reproduced. For instance, for $S=1/2$ and an initial product state one can show that (\ref{eq:lengthconstr_broken}) leads to $\sum_{\alpha}\braket{\widetilde{S}_i^{\alpha}(0)^2} = 1 \neq \frac{3}{4}=S(S+1)$.
Furthermore, we note that this quantity is related to the Schwinger boson constraint squared by
\begin{equation}
\braket{\hat{n}_i^2} = 4 \braket{\hat{\vec{S}}_i^2}-2\braket{\hat{n}_i} . \numberthis
\end{equation}
Since $\braket{\hat{n}}=2S$, this shows that the second-order identity $\braket{\hat{n}^2}=(2S)^2$ is not fulfilled in our approximation.

Our 2PI approach involves two approximations: the Gaussian approximation of the initial conditions and the $1/N$ expansion to NLO of the effective action. The latter, however, has in our case no influence on the value of $\braket{\hat{n}_i^2}$ and $\braket{\hat{\vec{S}}_i^2}$ at the initial time. To understand why, note that all higher-order diagrams beyond NLO in $1/N$ involve more than two interaction vertices. Their contribution to the self-energy $\Pi$, which is obtained from Eq.~(\ref{4_D_SE}), will thus involve at least one memory integral. Such integrals vanish at initial time and hence the NLO expression for $\Pi$ becomes exact at $t=0$.

The reason for the violation of the identities $\braket{\hat{n}_i^2}=(2S)^2$ and $\braket{\hat{\vec{S}}_i^2} = S(S+1)$ lies, therefore, in the Gaussian approximation of the initial non-Gaussian Fock state. This should come as no surprise since $\braket{\hat{n}_i^2}$ and $\braket{\hat{\vec{S}}_i^2}$ involve four-point expectation values in the $\varphi$ variables, which are obviously not captured in a Gaussian approximation.

Non-vanishing initial four-point, or, more generally, $n$-point functions can be taken into account by introducing additional $n$-point sources in the generating functional  which only have support at the initial time.
One possibility to treat such terms is to consider them as additional (time non-local) interaction vertices~\cite{Garny09}.
Another possibility is to use $nPI$ effective actions~\cite{Berges20042} which lead to self-consistent equations for $n$-point correlators.
\section{Energy}
\label{app:energy}
The conservation of energy is guaranteed due to the fact that all 2PI approximations fulfill global conservation laws~\cite{Baym1962}.

To explicitly compute the energy, we start by noting that it is the conserved charge corresponding to time translation invariance. It can therefore be deduced from the $2$PI effective action by calculating its change under the transformation $t\rightarrow t+\epsilon(t)$ and writing it as~\cite{arrizabalaga_equilibration_2005}
\begin{equation}
\delta\Gamma[G,D]=\int_\mathcal{C}\mathrm{d}t E(t)\partial_t\epsilon(t).
\end{equation}
A partial integration together with the fact that the variation of $\Gamma[G,D]$ vanishes for solutions of the equations of motion [see Eq.~(\ref{eq:eomeffaction})] directly proves the conservation of the energy function E.

The explicit calculation closely follows the one for the $O(N)$ model~\cite{arrizabalaga_equilibration_2005}, and yields
\begin{align}
E&=\frac{1}{8}\sum_{i,\gamma} \left(\bar{\chi}^\gamma_i(t)+2B_i^\gamma\right)\mathcal{K}^\gamma_{ab}F^{ab}_{ii}(t,t)\notag\\&\quad+\frac{1}{2}\sum_{ij,\gamma}J_{ij}^\gamma \hat D^{F,\gamma\gamma}_{ij}(t,t).
\label{eq_Energy}
\end{align}
This is in fact equal to taking the expectation value of the Hamiltonian [Eq.~(\ref{eq:Hgeneral_spins})] and inserting the expressions for the magnetization [Eq.~(\ref{eq:spin1pt_phichi})] and the spin-spin connected correlator [Eq.~(\ref{eq:spin2ptF_phichi})] in terms of Schwinger boson and auxiliary field correlators. The first term in Eq.~(\ref{eq:spin1pt_phichi}) arises due to the mean field contribution and the magnetic field and the second term due to the quantum fluctuations. 

Since the initial states considered in this work correspond to product states, the energy at initial time is entirely determined by the magnetizations, i.e. by the mean-field and magnetic field contributions (first term in Eq.~(\ref{eq_Energy})). Within the Gaussian approximation employed for the initial state, the magnetizations are set to the correct value at initial time and, thus, the value of the energy in our 2PI approach agrees with the value of the energy in the exact theory at the initial time, and hence at all later times. In App.~\ref{Num_impl} we comment on the numerical conservation of the energy as given by Eq.~(\ref{eq_Energy}).

We note that a slightly different approach to obtaining the energy is given in Ref.~[\citen{Weidinger2017}], where the Heisenberg equations of motion are used to express four-point-functions of Schwinger bosons in terms of $G$ and $D$. Following their approach yields in our case
\begin{align}
E &= -\frac{1}{4}\sum_i\left[\mathbb{1}\otimes i\sigma_y\right]^{ab}\partial_t F^{ba}_{ii}(t,t')\bigg|_{t=t'}\notag\\
&\quad+\frac{1}{8}\sum_{i,\gamma} B_i^\gamma\mathcal{K}^\gamma_{ab}F^{ab}_{ii}(t,t) .
\end{align}
One can analytically show that both expressions are equal by inserting Eq.~(\ref{eq:eomF_NLO}) and the Schwinger Boson self energies in Eqs.~(\ref{4_Sigma_F}),(\ref{4_Sigma_rho}). Recombining the resulting terms by employing the equations of motion for the auxiliary field correlator [Eq.(\ref{eq:eomDF_NLO})] then yields Eq.~(\ref{eq_Energy}).
\section{Numerical Implementation}
\label{Num_impl}
The dynamical equations as given in Eqs.~(\ref{eq:eomF_NLO}) and~(\ref{eq:eomrho_NLO}) are first order integro-differential equations which are hard to solve analytically, especially when supplemented with additional Volterra-type integral equations for the self-energy as in the 1/N expansion. Therefore, we revert to a numerical evaluation of these expressions.

For a rather old but well referenced review of numerical techniques for general Volterra type integral and integro-differential equations see Ref.~[\citen{Brunner82}]. The quite sophisticated methods in use for evaluation in thermal equilibrium (which requires solving a  self-consistency equation iteratively rather than propagating an initial value problem) are described in Ref.~[\citen{Balzer11}].

The numerical techniques which have been primarily used to solve the dynamical equations for non-equilibrium problems mostly use first an extended Newton-Cotes discretisation to compute the integrals and then use the result to solve the differential part of the equation with standard differential equation solvers. One complication arises from the requirement to fulfill conservation laws in the evaluation, in our case the energy, total magnetization (in some models) and the Schwinger boson constraint. For relativistic field theories, a very elegant solution of this requirement lies in the symmetric discretisation of the second-order time derivative as described in Ref.~[\citen{Garny2009}] and in Ref.~[\citen{Berges:2002wr}] with additional fermionic fields. For non-relativistic theories, this is not possible, as the symmetric derivative is a second-order discretisation of a first-order derivative, which is inherently unstable~\cite{numrec07}, as we explicitly checked in our case.
The most widely used method to circumvent this complication has been the usage of predictor-corrector algorithms, which will be our method of choice in this work and will be explained in the following. See e.g.~Ref.~[\citen{Bock2016}] for a higher-order method, or Ref.~[\citen{Sexty11}] for a neat trick to separate out the free evolution part.

The next sections are organized as follows. First, we will introduce the predictor corrector method. Then, for the evaluation of the right hand sides of the differential equations, we show how to discretize the memory integrals and how the order of the evaluation of the various quantities proceeds. We end by discussing tricks to lower the resource consumption and how conservation laws are fulfilled with our method.

\subsection{Predictor-corrector method}
Below, we denote the discretised right hand sides of the equations for $F$ and $\rho$, Eqs.~(\ref{eq:eomF_NLO}), (\ref{eq:eomrho_NLO}), with $\mathrm{RHS}^F$ and $\mathrm{RHS}^\rho$.

For the diagonal steps we will furthermore need the evolution equation for the second time argument of $F$,
\begin{align}
	&\partial_{t_2}F_{ii}^{ab}(t_1,t_2)=\notag\\ &i\left[\mathbb{1}\otimes\sigma_y\right]^{bc}\bigg\lbrace\frac{1}{2} \sum_{\gamma} \big( \bar{\chi}_i^{\gamma}(t_2) + B_i^\gamma \big) \mathcal{K}_{cd}^{\gamma}F_{ii}^{ad}(t_1,t_2)+ \nonumber\\
&\int_0^{t_1}\mathrm{d}t\,\rho^{ad}_{ii}(t_1,t)\Sigma^{F,dc}_{ii}(t,t_2)-\int_0^{t_2}\mathrm{d}t\,F_{ii}^{ad}(t_1,t)\Sigma_{ii}^{\rho,dc}(t,t_2)\bigg\rbrace.
\end{align}
which follows from Eq.~\ref{eq:eomF_NLO} by exchanging $t_2\leftrightarrow t_1$ and using the symmetry properties of $F$. We denote the discretized RHS of above equation with RHST2 below.

For a single time step from $t_1$ to $t_1+1$, the predictor-corrector algorithm proceeds as:
\begin{enumerate}
\item\textbf{Predict $F(t_1+1,t_2)$ and $\rho(t_1+1,t_2)$ for $t_2\leq t_1$}
\begin{equation}
F^{ab}_{ii}(t_1+1,t_2) = F^{ab}_{ii}(t_1,t_2) + \Delta t\times \mathrm{RHS}^{F,ab}_{ii}(t_1,t_2),
\end{equation}
\begin{equation}
\rho^{ab}_{ii}(t_1+1,t_2) = \rho^{ab}_{ii}(t_1,t_2) + \Delta t\times \mathrm{RHS}^{\rho,ab}_{ii}(t_1,t_2).
\end{equation}
\item\textbf{Predict $F(t_1+1,t_1+1)$ and set $\rho(t_1+1,t_1+1)$ to equal-time commutation relations.}
\begin{align}
&F^{ab}_{ii}(t_1+1,t_1+1)=F^{ab}_{ii}(t_1,t_2)+\notag\\
&\qquad \Delta t\times \big(\mathrm{RHS}^{F,ab}_{ii}(t_1,t_1)+\mathrm{RHST2}^{F,ab}_{ii}(t_1,t_1)\big),
\end{align}
\begin{equation}
\rho^{ab}_{ii}(t_1+1,t_1+1)=\rho^{ab}_{ii}(0,0).
\end{equation}
\item\textbf{Evaluate the RHSs for $t_2\leq t_1+1$(for details see below in subsection 2. and 3.)} \begin{align}
&\mathrm{RHS}^{F,ab}_{ii}(t_1+1,t_2),\\
&\mathrm{RHS}^{\rho,ab}_{ii}(t_1+1,t_2),\\
&\mathrm{RHST2}^{F,ab}_{ii}(t_1+1,t_1+1).
\end{align}
\item\textbf{Correct $F$ and $\rho$ for $t_2\leq t_1$ }
\begin{align}
F^{ab}_{ii}(t_1+1,t_2) &= F^{ab}_{ii}(t_1,t_2) + \frac{\Delta t}{2}\Big(\mathrm{RHS}^{F,ab}_{ii}(t_1+1,t_2)\nonumber\\&\qquad+\mathrm{RHS}^{F,ab}_{ii}(t_1,t_2)\Big),\\
\rho^{ab}_{ii}(t_1+1,t_2) &= \rho^{ab}_{ii}(t_1,t_2) + \frac{\Delta t}{2}\Big(\mathrm{RHS}^{\rho,ab}_{ii}(t_1+1,t_2)\nonumber\\&\qquad+\mathrm{RHS}^{\rho,ab}_{ii}(t_1,t_2)\Big).
\end{align}
\item\textbf{Correct $F(t_1+1,t_1+1)$.}
\begin{align}
&F^{ab}_{ii}(t_1+1,t_1+1)=F^{ab}_{ii}(t_1,t_2)+\notag\\
& \Delta t\times \big(\mathrm{RHS}^{F,ab}_{ii}(t_1,t_1) +\mathrm{RHST2}^{F,ab}_{ii}(t_1,t_1) +\notag\\&\mathrm{RHS}^{F,ab}_{ii}(t_1+1,t_1+1)+\mathrm{RHST2}^{F,ab}_{ii}(t_1+1,t_1+1)\big),
\end{align}
\end{enumerate}
The steps 3.-5. are then iterated up to a certain convergence, which improves the fulfilment of conservation laws (see below). Furthermore, the evaluated RHSs can then be stored for the predictor step in the next timestep.
\subsection{Discretization of memory integrals}

We discretize all memory integrals with the trapezoidal rule, such that the RHSs become
\begin{align*}
&\mathrm{RHS}^{F,ab}_{ii}(t_1,t_2) \\
&= \left[\mathbb{1}\otimes i\sigma_y\right]^{ac}\bigg\lbrace\frac{1}{2}\sum_{\gamma}(\bar{\chi}^{\gamma}_{i}(t_1)+B_i^{\gamma})\mathcal{K}_{cd}^{\gamma}F^{db}_{ii}(t_1,t_2)\\
&+ \frac{\Delta t}{2} \bigg(\Sigma^{\rho,cd}_{ii}(t_1,0)F^{db}_{ii}(0,t_2)+\Sigma^{\rho,cd}_{ii}(t_1,t_1)F^{db}_{ii}(t_1,t_2)\bigg)\\
&+\Delta t\sum_{l=1}^{t_1-1} \Sigma^{\rho,cd}_{ii}(t_1,l)F^{db}_{ii}(l,t_2)\\
&-\frac{\Delta t}{2} \bigg(\Sigma^{F,cd}_{ii}(t_1,0)\rho^{db}_{ii}(0,t_2)+\Sigma^{F,cd}_{ii}(t_1,t_2)\rho^{db}_{ii}(t_2,t_2)\bigg)\\
&-\Delta t\sum_{l=1}^{t_2-1} \Sigma^{F,cd}_{ii}(t_1,l)\rho^{db}_{ii}(l,t_2)\bigg\rbrace \numberthis,
\label{eq:discreteRHSF}
\end{align*}
\begin{align*}
&\mathrm{RHS}^{\rho,ab}_{ii}(t_1,t_2) \\
&= \left[\mathbb{1}\otimes i\sigma_y\right]^{ac}\bigg\lbrace\frac{1}{2}\sum_{\gamma}(\bar{\chi}^{\gamma}_{i}(t_1)+B_i^{\gamma})\mathcal{K}_{cd}^{\gamma}\rho^{db}_{ii}(t_1,t_2)\\
&+ \frac{\Delta t}{2} \bigg(\Sigma^{\rho,cd}_{ii}(t_1,t_2)\rho^{db}_{ii}(t_2,t_2)+\Sigma^{\rho,cd}_{ii}(t_1,t_1)\rho^{db}_{ii}(t_1,t_2)\bigg)\\
&+\Delta t\sum_{l=t_2+1}^{t_1-1} \Sigma^{\rho,cd}_{ii}(t_1,l)\rho^{db}_{ii}(l,t_2)\bigg\rbrace. \numberthis
\label{eq:discreteRHSrho}
\end{align*}

The equations for $\hat D^F$ ($\hat D^\rho$) are not explicit with the trapezoidal rule, i.e. the RHS depends on the quantity to be determined. This can however be circumvented by a simple matrix inversion in the auxiliary field and space indices, such that Eqs.~(\ref{eq:eomDF_NLO}) and (\ref{eq:eomDrho_NLO}) become
\begin{align*}
\label{4_D_F}
&\hat{D}^{F,\epsilon\beta}_{ij}(t_1,t_2) = \bigg[ \mathbb{1}-\frac{\Delta t}{2}\Pi^\rho(t_1,t_1)J\bigg]^{-1,\epsilon\alpha}_{ik}\times\\&\bigg[-\Pi_{kk}^{F,\alpha\beta}(t_1,t_2)\delta_{kj}+\\&\quad\Delta t\sum_{m, \delta}J_{km}^{\delta}\bigg\lbrace\frac{1}{2}\Pi_{kk}^{\rho, \alpha\delta}(t_1,0)\hat{D}^{F,\beta\delta}_{jm}(t_2,0)\\
&\qquad+\frac{1}{2}\Pi_{kk}^{F,\alpha\delta}(t_1,0)\hat{D}^{\rho,\beta\delta}_{jm}(t_2,0)\\&\qquad+\frac{1}{2}\Pi_{kk}^{F,\alpha\delta}(t_1,t_2)\hat{D}^{\rho,\delta\beta}_{mj}(t_2,t_2)\\
&\qquad+\sum_{l=1}^{t_1-1} \Pi_{kk}^{\rho, \alpha\delta}(t_1,l)\hat{D}^{F,\delta\beta}_{mj}(l,t_2)\\&\qquad+\sum_{l=1}^{t_2-1} \Pi_{kk}^{F,\alpha\delta}(t_1,l)\hat{D}^{\rho,\beta\delta}_{jm}(t_2,l)\bigg\rbrace\bigg], \numberthis
\end{align*}
\begin{align*}
\label{4_D_rho}
&\hat{D}^{\rho,\alpha\beta}_{kj}(t_1,t_2) = \bigg[ \mathbb{1}-\frac{\Delta t}{2}\Pi^\rho(t_1,t_1)J\bigg]^{-1,\epsilon\alpha}_{ik}\times\\&\bigg[-\Pi_{kk}^{\rho,\alpha\beta}(t_1,t_2)\delta_{kj}+\\&\quad+\frac{\Delta t}{2}\sum_{m, \delta} \Pi_{kk}^{\rho,\alpha\delta}(t_1,t_2)J_{km}^{\delta}\hat{D}^{\rho,\delta\beta}_{mj}(t_2,t_2)+\\
&\quad
+\Delta t\sum_{m, \delta} \sum_{l=t_2+1}^{t_1-1} \Pi_{kk}^{\rho,\alpha\delta}(t_1,l)J_{km}^{\delta}\hat{D}^{\rho,\delta\beta}_{mj}(l,t_2).\bigg] \numberthis
\end{align*}
All the other expressions, especially the self energies $\Pi,\Sigma$ are given by simple multiplications or additions. 
\subsection{Order of evaluation}

In order to ensure that each evaluation step only requires known quantities, the following procedure is followed when evaluating the $\mathrm{RHS}(t_1,t_2)$ for $t_1\leq t_2$ (i.e. step 3. in the predictor-corrector scheme):
\begin{enumerate}
\item Calculate all $\Pi^{\rho}(t_1,t_2)$, $\Pi^{F}(t_1,t_2)$ and $\bar{\chi} (t_1)$ for $t_2 \leq t_1$ from Eqs.~(\ref{4_Pi_F}), (\ref{4_Pi_rho}) and (\ref{eq:eomChi_NLO}).  
\item Calculate $D^\rho (t_1,t_2)$ for $t_2 \leq t_1$ from Eq.~(\ref{4_D_rho}). Note that for $t_1=t_2$, the memory integrals do not contribute.
\item Calculate $D^F (t_1,t_2)$ for $t_2 < t_1$ from Eq.~(\ref{4_D_F}).
\item Calculate $D^F (t_1,t_2)$ for $t_2 = t_1$ from Eq.~(\ref{4_D_F}). Note that the memory integrals depend on $D^F (t_1,t_2)$ with $t_2 < t_1$.
\item Calculate $\Sigma^F (t_1,t_2)$ and $\Sigma^\rho (t_1,t_2)$ for $t_2 \leq t_1$ from Eqs.~(\ref{4_Sigma_F}), (\ref{4_Sigma_rho}).
\item Calculate $\mathrm{RHS}^\rho$ from Eq.~(\ref{eq:discreteRHSrho}). 
\item Calculate $\mathrm{RHS}^F$ from Eq.~(\ref{eq:discreteRHSF}).
\end{enumerate}

\subsection{Conservation laws\label{Num_cons}}

The Schwinger boson constraint $\braket{\hat n_i}=2S$ computed as
\begin{align}
\braket{\hat n_i} &= \frac{1}{2}\left(\sum_a F^{aa}_{ii}(t,t)-2\right),
\end{align}
is conserved up to $10^{-15}$ relative error (i.e. machine precision) in our scheme. Note that this is not the case when using other ways of calculating the diagonal step (step 4. in predictor-corrector scheme), for example the one used in Ref.~[\citen{Rey04}].

Additionally, the total $S^z$ magnetization is conserved up to a similar precision as the constraint in models in which it is a conserved quantity such as the Heisenberg model.
The energy in all simulations presented in this work is at least conserved up to $10^{-3}$ relative error, and typically up to $10^{-6}$. We checked that the error decreases as $\Delta t^2$ with decreasing time step. When employing a memory cut in section \ref{ch_rel_qu}, we observed a slight drift in the energy due to the explicit dependence of the energy on the memory integrals as visible in Fig.~\ref{fig_energy} for late times. This drift is on the order of $10^{-3}$ at the latest time considered, decreases with increasing memory zone and we checked that it did not have any sizeable effects on the results presented. Furthermore, we checked that both methods of calculating the energy derived in App.~\ref{app:energy} coincide numerically if the evaluate-correct step is sufficiently iterated.

\subsection{Resource consumption and possible improvements}
\label{ch_ressources}
The evaluation of the Kadanoff-Baym equations is very costly, especially with regards to memory consumption. This is due to the necessity to store the whole past for the evaluation of the memory integrals. The objects which, in our case, overshadow all others in terms of memory requirement are $D^F$ and $D^\rho$ as they are functions of two lattice site indices. Taking into account that both of them contain $9$ components, storing each of them requires a priori around $670\,\mathrm{GiB} $ of memory for $1000$ time steps, $100$ lattice sites and double precision. In this estimate, a factor of $1/2$ was already saved by only saving the lower triangle of the $t_1,t_2$ plane of all correlation functions, as the upper triangle can be deduced due to their (anti-)symmetry.
\begin{figure}[htp]
\centering
\includegraphics{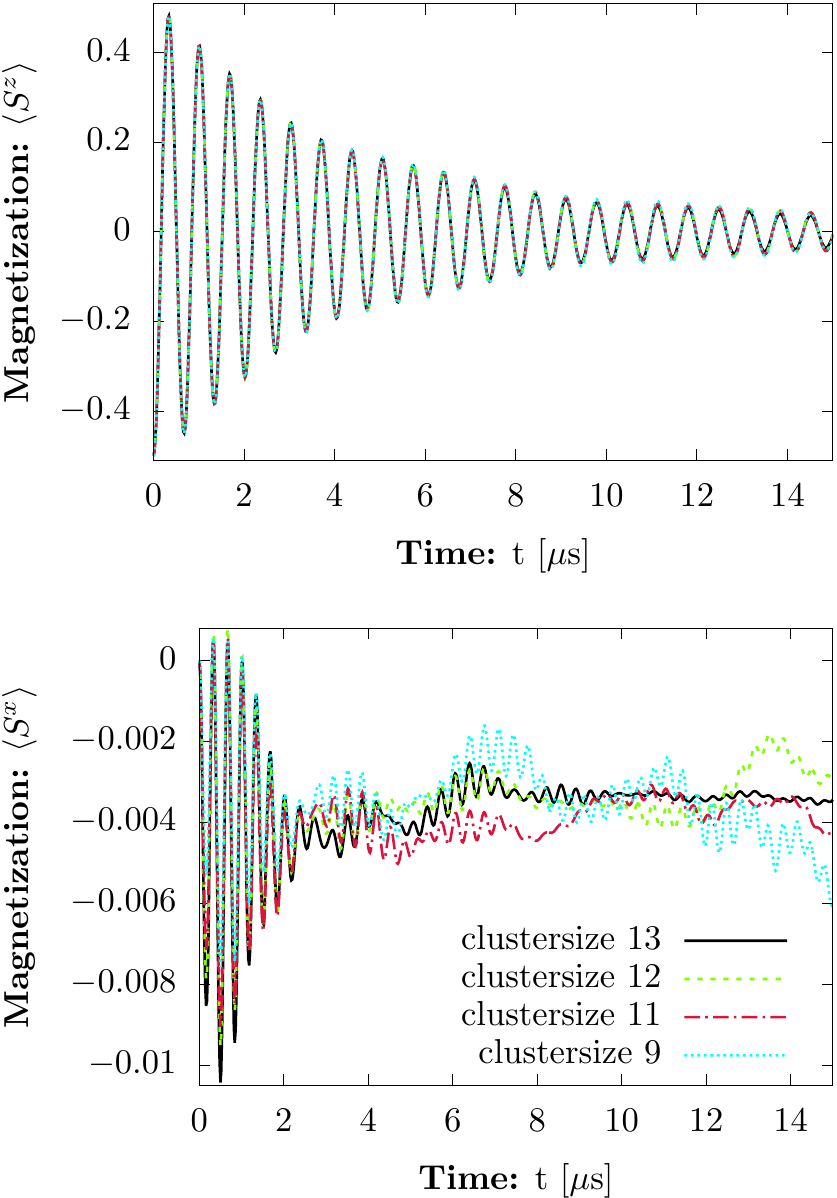}
\caption{Comparison of different cluster sizes in MACE for the plots shown in Fig. \ref{fig_mag_N100_2}. Note that all cluster sizes lie on top of each other in the plot of the $S^z$ component.}
\label{fig_MACE_convergence}
\end{figure}

A large increase in efficiency in both execution time and memory can be achieved when dealing with a system which forgets its initial state at a sufficiently fast rate, as, for example, encountered in a thermalizing or relaxing system. In this case, one can neglect at late times the parts of the memory integrals which involve times in the remote past. This means that one only needs to store the times closest to the current time and gradually shift the memory zone forwards. This procedure is nicely described in Ref.~[\citen{Garny2009}]. The truncation of the memory integrals makes it, in principle, possible to evolve the system to arbitrarily late times using a fixed amount of memory and within a computational time which scales, asymptotically, as a \emph{linear} function of evolution time. The latest time achievable may be, however, bounded by the growth of the error in the memory integral truncation.

The memory consumption can furthermore be reduced by a factor of two by using the (anti-)symmetry properties of $\hat D^F$ and $\hat D^\rho$ such that only half of the array needs to be stored.

In the example shown in Fig.~\ref{fig_mag_N100_2}, this procedure reduces the total memory consumption from $\approx 1\,\mathrm{TiB}$ to a mere $17\,\mathrm{GiB}$. The single realization shown in Fig.~\ref{fig_mag_N100_2} took approximately one week of computation time on a cluster with $64$ cores of type Intel Xeon Phi 7210-F and a time step of $\Delta t = 0.01$. For the considerably smaller systems studied in chapters \ref{MBL} and \ref{Rel_dyn_XXZ} the computation time was on the order of $30$ minutes for a single disorder realization and a time step of $\Delta t = 0.02$, but without memory cut.

Lastly, we comment on different numerical methods to solve the Kadanoff-Baym equations in our case. While a symmetric time derivative does not need a sophisticated predictor-corrector scheme to fulfil conservation laws, it proved to be inherently unstable due to it being a second-order discretization of a first-order time derivative. A simple Riemann discretization of the memory integrals in the equations for $D^{\rho/F}$, while circumventing the matrix inversion needed with the trapezoidal rule, lead to a substantially larger error in the energy. We expect higher-order predictor-corrector methods as used in Ref.~[\citen{Bock2016}] to lead to better convergence properties, which may improve the area of application of the present method.
\subsection{MACE and its convergence properties\label{MACE}}

The MACE method \cite{Hazzard14} consists in taking a spin $i$ in the system, building a cluster around it composed of its closest neighbors, and solving the cluster dynamics by exact diagonalization to compute the magnetization dynamics $\braket{S_i^\alpha}(t)$ of the spin $i$. One may also build the cluster by choosing the spins $j$ with the largest couplings $|J_{ij}|$, which we checked leads, in our case, to the same result. To obtain the total magnetization, this procedure is repeated with each spin of the ensemble and then averaged over. We estimate that MACE has reached approximate convergence when increasing the cluster size does not lead to appreciable differences. MACE has been previously employed to describe the magnetization dynamics of spin systems with quenched disorder in cold dipolar molecules \cite{Hazzard14, Yan2013} or Rydberg atoms \cite{Signoles17}, showing convergence for cluster sizes of around $10-12$ spins. While modifications of the method for applications in lattice spin systems have been proposed \cite{White2017}, computation of other observables such as two-point functions remains challenging.

Fig.~\ref{fig_MACE_convergence} shows the evolution of the magnetizations displayed in Fig.~\ref{fig_mag_N100_2} as obtained from MACE for different cluster sizes.
While the results for the $S^z$ component from all cluster sizes shown lie on top of each other, this is not the case for the $S^x$ component. Even at rather early times the curves for $S^x$ show significant deviations up to the largest size considered, without any sign of convergence. At longer times all cluster sizes seem to converge to a negative value for the magnetization, but the final value shows considerable fluctuations between cluster sizes.
Because of this we conclude that the MACE result for $\braket{S^z}$ has converged, whereas $\braket{S^x}$ has not.

\bibliography{Bibliography}

\end{document}